# Solution of the Schrodinger equation using tridiagonal representation approach in nonrelativistic quantum mechanics: a Pedagogical Approach

E.O. Oghre[1], T. J. Taiwo[1], and A.N. Njah[2]

[1]*Department of Mathematics, University of Benin, Benin City, Edo State 300283, Nigeria*
[2]*Department of Physics, University of Lagos, Akoka. Lagos State 101017, Nigeria*

**Abstract**: We present the pedagogic method of tridiagonal representation approach algebric method for the solution of Schrodinger equation in nonrelativistic quantum mechanics for conventional potential functions. However, we solved a new three parameters potential function (100) recently encountered and using it as an example.

## 1. Introduction

Exactly solvable potentials have been the focus of research since the beginning of nonrelativistic quantum mechanics. The essences of these are based on the benefits derived from getting the associated properties of the underlying physical systems such as: eigenvalues, eigenfunctions, scattering phase shift etc. Also these potentials constitute good effective models for physical system or approximations thereof. Efforts to classify exactly solvable systems have been carried out by different scientists. For instance, Infeld and Hull [2] have succeeded in classifying most of the solvable Hamiltonians based on the Darboux factorization method [1]. Further researches were done by Natanzon, Gendenstein and others [3-4]

As a result, solvable potentials have been classified into exactly solvable, conditionally exactly solvable or quasi exactly solvable. In those developments, the main aim is to find the solutions of the energy eigenvalues wave equation $H|\psi\rangle = E|\psi\rangle$, where $H$ is the Hamiltonian and $E$ is the energy which is either discrete (for bound states) or continuous (for scattering states). In most of these studies, the matrix representation of the Hamiltonian is written in the bound states of the associated potential resulting in a diagonal structure exhibiting the eigenvalues or the energy spectrum.

Here, we present the Tridiagonal Representation Approach (TRA). The Hamiltonian operator (wave operator) diagonal matrix representation restriction is relaxed but required to be tridiagonal and symmetric. This will therefore require expanding the solution space of the wave operator in a basis set that does not have to be an element of the eigen states of the Hamiltonian but must be complete and square integrable with respect to an appropriate configuration space measure. Applying the TRA, we have been able to get new solvable potential functions (but are not solvable or known in conventional quantum mechanics using classical methods) including the well-known ones.

Having used the TRA for some time now with remarkable success [6-8]; we deem it necessary to introduce this algebric approach to students of physics, mathematics, chemistry and engineering. Here, we restrict the treatment to nonrelativistic quantum mechanics.
We present the TRA using a Pedagogical approach suitable for the classroom. Section 2, contains introduction to the TRA – theoretical formulation. In section 3, the idea of space transformation and basis set will be presented. In section 4, we introduce the classical orthogonal polynomials and their properties. In section 5, we show how to apply the TRA in the solution of a given problem.

## 2. Theoretical formulation of the TRA

The conventional theory of quantum mechanics of a close system deals essentially with following time-independent eigenvalue problem:

$$\hat{H}|\psi\rangle = E|\psi\rangle \tag{1}$$

where $\hat{H}$ is the Hamiltonian of the system. This equation is general and does not depend on any coordinate system or representation. *For solving this equation, we need to represent it in a given basis system and their lies the complexity in solving the equation.* Specifically, how do we represent the Hamiltonian operator assuming that its Hilbert space could be discretized?

Now consider a discrete, complete (finite or infinite), and square integrable basis which is made of kets $|\phi_1\rangle, |\phi_2\rangle, |\phi_3\rangle, ..., |\phi_n\rangle$ denoted by $\{|\phi_n\rangle\}$. For simplicity in this introduction to the TRA, we assume that the basis are real and their overlap matrix $\langle\phi_n|\phi_m\rangle$ is tridiagonal and symmetric. The orthonormality condition of the basis is expressed as $\langle\phi_n|\phi_m\rangle = \delta_{nm}$, where $\delta_{nm}$ is the *Kronecker delta* symbol defined by

$$\delta_{nm} = \begin{cases} 1, & n = m, \\ 0, & n \neq m. \end{cases} \tag{2}$$

The completeness of the basis is given as $\sum_{n=1}^{\infty}|\phi_n\rangle\langle\phi_m| = \hat{I}$, where $\hat{I}$ is a unit vector. Inserting the unit vector between $\hat{H}$ and $|\psi\rangle$ in (1) and multiplying by $\langle\phi_m|$ gives

$$\langle\phi_m|\hat{H}\left(\sum_n|\phi_n\rangle\langle\phi_m|\right)|\psi\rangle = E\langle\phi_m|\left(\sum_n|\phi_n\rangle\langle\phi_m|\right)|\psi\rangle \tag{3}$$

which is same as

$$\sum_n H_{nm}\langle\phi_n|\psi\rangle = E\sum_n\langle\phi_n|\psi\rangle\delta_{nm} \tag{4}$$

where $H_{nm} = \langle\phi_n|\hat{H}|\phi_m\rangle$, the Hamiltonian matrix elements, and $\langle\phi_n|\phi_m\rangle = \delta_{nm}$. Equation (4), can be written as

$$\sum_n [H_{nm} - E\delta_{nm}]\langle\phi_n|\psi\rangle = 0 \tag{5}$$

This equation stands for an infinite, homogenous system of equations for the coefficients $\langle\phi_n|\psi\rangle$. Therefore there exist nonzero solutions if and only if the determinant $Det(H_{nm} - E\delta_{nm}) = 0$ vanishes. That is

$$\begin{vmatrix} H_{11} - E & H_{12} & H_{13} & \ldots & H_{1N} \\ H_{21} & H_{22} - E & H_{23} & \ldots & H_{2N} \\ H_{31} & H_{32} & H_{33} - E & \ldots & H_{3N} \\ \vdots & \vdots & \vdots & \ddots & \vdots \\ H_{N1} & H_{N2} & H_{N3} & \ldots & H_{NN} - E \end{vmatrix} = 0 \tag{6}$$

Equation (6) is an Nth order equation in *E*; its solutions will yield the energy spectrum of the quantum system: $E_1, E_2, E_3, ..., E_N$. With these values, one can easily get the corresponding set of eigenvectors $|\phi_1\rangle, |\phi_2\rangle, ..., |\phi_N\rangle$. Since our basis system is a set of discrete, complete, and orthonormal elements using (2); equation (6) becomes

$$\begin{vmatrix} H_{11} - E & 0 & 0 & \ldots & 0 \\ 0 & H_{22} - E & 0 & \ldots & 0 \\ 0 & 0 & H_{33} - E & \ldots & 0 \\ \vdots & \vdots & \vdots & \ddots & \vdots \\ 0 & 0 & 0 & \ldots & H_{NN} - E \end{vmatrix} = 0 \tag{7}$$

This is a diagonalization of the Hamiltonian matrix $H_{nm} = \langle\phi_n|\hat{H}|\phi_m\rangle$, of the quantum system which will yield the energy spectrum as well as the state vectors. From linear algebra knowledge, one can easily say equation (1) can be theoretically written as (after the Hamiltonian operator has acted on the basis element)

$$\begin{vmatrix} H_{11} & 0 & 0 & \ldots & 0 \\ 0 & H_{22} & 0 & \ldots & 0 \\ 0 & 0 & H_{33} & \ldots & 0 \\ \vdots & \vdots & \vdots & \ddots & \vdots \\ 0 & 0 & 0 & \ldots & H_{NN} \end{vmatrix} |\psi\rangle = \begin{vmatrix} E_1 & 0 & 0 & \ldots & 0 \\ 0 & E_2 & 0 & \ldots & 0 \\ 0 & 0 & E_3 & \ldots & 0 \\ \vdots & \vdots & \vdots & \ddots & \vdots \\ 0 & 0 & 0 & \ldots & E_N \end{vmatrix} |\psi\rangle \tag{8}$$

the *Es*: $E_1, E_2, E_3, \ldots, E_N$ in equation (8) are the eigenvalues of the Hamiltonian matrix $H_{nm}$ which give the energy spectrum of the quantum system and which is why a quantum system is easily defined in the form of equation (1).

*Now, the question is what made it possible for the Hamiltonian matrix to become diagonalized? The answer to is the basis elements (basis set).* The Hamiltonian matrix becomes diagonal due to the discrete, completeness, and orthonormalization of the basis set. These properties make this basis set a countable infinite element of the Hilbert space. So, the basis set determines it all.

However, in the TRA, we tend to choose a basis set with element that will make the Hamiltonian operator $\hat{H}$ and/or the Wave operator $\hat{J} = (\hat{H} - E)$ (matrix elements) become a tridiagonal and symmetric matrix. That is for equation (7) to become

$$\begin{vmatrix} H_{11} - E & H_{21} & 0 & \ldots & 0 \\ H_{21} & H_{22} - E & H_{32} & \ldots & 0 \\ 0 & H_{32} & H_{33} - E & \ddots & 0 \\ \vdots & \vdots & \ddots & \ddots & H_{N,N-1} \\ 0 & 0 & 0 & H_{N,N-1} & H_{NN} - E \end{vmatrix} = 0 \tag{9}$$

for easy display, we rewritten equation (9) as

$$\begin{pmatrix} a_{11} & b_{21} & 0 & \ldots & 0 \\ b_{21} & a_{22} & b_{32} & \ldots & 0 \\ 0 & b_{32} & a_{33} & \ddots & 0 \\ \vdots & \vdots & \ddots & \ddots & b_{N,N-1} \\ 0 & 0 & 0 & b_{N,N-1} & a_{NN} \end{pmatrix} \tag{10}$$

In the TRA, the basis sets that will make the wave operator (Hamiltonian operator) matrix representation become tridiagonal and symmetric are the special functions (classical orthogonal polynomials): Hermite, Legendre, Jacobi, Laguerre and so on. These functions have interesting properties that will result in the tridiagonal – symmetric form of the wave operator (Hamiltonian operator). Such properties include: weight function, nature of generating function, orthogonality, distribution and density of the polynomial zeros, recursion relation, asymptotics, and differential or difference equation. Also, they are infinite discrete square – integrable functions and hence belong to the Hilbert space. As a result of these, in the TRA, our basis set $\{|\phi_n\rangle\}$ will have elements $\{\phi_n\}_{n=0}^{\infty}$ written in term of orthogonal polynomials. Achieving (9) in the TRA, will enlarge the solution space of the wave operator (Hamiltonian operator) of quantum system to incorporate new exactly solvable potentials. In fact, in the TRA, we can return back to the diagonal restriction place on wave operator (Hamiltonian operator) which made it possible to get conventional exactly solvable potentials. All these features will be shown later. In quantum mechanics class, we are introduced to the one dimensional time- independent Schrodinger equation for a microscopic particle of mass *m* in the field of a one dimensional time – independent potential function $V(x)$. This equation is given as

$$\left[ -\frac{\hbar^2}{2m}\frac{d^2}{dx^2} + V(x) \right] \psi(x, E) = E\psi(x, E); \quad x \in R \tag{11}$$

where $E$ is the particle's energy and $\psi(x, E)$ is the wavefunction that describes the state. The physical configuration space coordinate belongs to the interval $x \in [x_-, x_+]$, which could be finite, infinite, or semi – infinite. Conventionally, we need to specify the potential function $V(x)$ as well as the boundary conditions, which can be obtained from the physical requirements of the system, in order to solve (11) and get our total solution as $\Psi(x,t) = \psi(x)e^{-iEt/\hbar}$. Based on the mathematical rigor in obtaining these solutions with the associated properties (energy spectrum, scattering phase shift, resonance, transmission coefficient, and reflection coefficient) as resulted in the classification of the quantum systems based on their associated potential functions as: *exactly – solvable potentials, quasi-exactly solvable potentials and conditional solvable potentials.*

However, in the TRA there are two possibilities[3]:

- *We might not need to insert the potential function $V(x)$ in the Schrodinger equation and yet based on specific coordinate space transformation that is compatible with our basis element; we can recover the potential function $V(x)$, energy spectrum formula, scattering phase shift, resonance and the wavefunction.*
- *We can also insert the potential function $V(x)$, do a specific space transformation compatible that is compatible with our basis element; and get the energy spectrum, scattering phase shift, resonance and the wavefunction.*

From the above points, one can infer the power of the basis element $\{\phi_n\}_{n=0}^{\infty}$. The different coordinate space transformations $y(x)$ compatible with the basis element will result in known conventional exact solvable potential functions (possibly modified versions of them), and new ones not known in the physics literature. Invariably, we can say for each known exactly-solvable potential functions in conventional quantum mechanics; there exists a specific coordinate space transformation $y(x)$ compatible with the basis element that will result in recovering each potential function and the association properties of the underlying quantum system.

That is advantage of the TRA. However, in either point, our basis element $\{\phi_n\}_{n=0}^{\infty}$ space configuration $y(x)$ has to be determined. *Hence, we normally transformed the Schrodinger equation to a new space configuration that is compatible with the basis element.* We show this in section 2. In formulating the TRA; we write the wavefunction as an infinite sum:

$$\psi(x,E) = \sum_n f_n(E) |\phi_n(x)\rangle \tag{12}$$

where $|\phi_n(x)\rangle$ is a set of square integrable discrete basis with element $\{\phi_n\}_{n=0}^{\infty}$ written in term of orthogonal polynomials. $x$ is the configuration space coordinate. The basis element is compatible with the domain of the Hamiltonian and satisfy the boundary conditions of the problem (usually, the vanishing of the wavefunction at the boundaries for bound states, whereas a sinusoidal behaviour at infinite boundaries for scattering states). Also, the expansion coefficients $f_n(E)$ are energy orthogonal polynomials which we will end up determining by comparism. The details of all these will be shown carefully later.

Equation (11) is same as (1) because $\hat{H} = -\frac{\hbar^2}{2m}\frac{d^2}{dx^2} + V(x)$, is the Hamiltonian operator. From (1), we have the

$$(\hat{H} - E)|\psi\rangle = 0 \tag{13}$$

where $\hat{J} = (\hat{H} - E)$ is the wave operator. In the TRA, we always require the wave operator to operate on the basis element to give

$$\langle \phi_n|(\hat{H} - E)|\phi_m\rangle = (a_n - E)\delta_{n,m} + b_n \delta_{n,m-1} + b_{n-1}\delta_{n,m+1} \tag{14}$$

where the real coefficients $\{a_n, b_n\}_{n=0}^{\infty}$ are functions of the physical parameters of the problem. Equation (14) is a tridiagonal-symmetric matrix as compared to equation (7). For instance, suppose we want to create matrix of order $5 \times 5$ starting with row and column named 0,1,2,3, and 4 ($Row: n = 0 \ 1 \ 2 \ 3 \ 4$ and $Column: m = 0 \ 1 \ 2 \ 3 \ 4$). Using (2), we start from the first row $n = 0$ and move across the column (subsequently repeating the procedure until we finally get to $n = 4$); then we will have the wave operator matrix representation as

$$J_{nm} = \begin{pmatrix} a_0 - E & b_0 & 0 & 0 & 0 \\ b_0 & a_1 - E & b_1 & 0 & 0 \\ 0 & b_1 & a_2 - E & b_2 & 0 \\ 0 & 0 & b_2 & a_3 - E & b_3 \\ 0 & 0 & 0 & b_3 & a_4 - E \end{pmatrix} \quad (15)$$

Here we see that the diagonal restriction of the wave operator has been lifted because of the basis element. So, the matrix wave equation in the TRA is obtained by using (2) in (13) and projecting on the left by $\langle \phi_m |$,

$$\sum_n f_n \langle \phi_m | \hat{H} | \phi_n \rangle = E \sum_n f_n \langle \phi_m | \phi_n \rangle$$
$$\sum_n H_{m,n} f_n = E \sum_n \Omega_{m,n} f_n \quad (16)$$

where $\Omega_{m,n} = \langle \phi_m | \phi_n \rangle$ is the overlap matrix of the basis elements which is the matrix representation of the identity[4]. So, the wave operator matrix representation is

$$J_{nm} = H_{nm} - E\Omega_{nm} \quad (17)$$

which we demand to be tridiagonal and symmetric and hence place restrictions on the types of potential functions we will get. Satisfying these restrictions will make the wave equation becomes

$$Ef_n = a_n f_n + b_{n-1} f_{n-1} + b_n f_{n+1} \quad (18)$$

This is the recursion relation that enables us to find the expansion coefficients in (12). So, our reference problem has translated into finding the solutions of the recursion relation for the expansion coefficients of the wave function. In most case, we get this by comparing to the recursion relation of well-known orthogonal polynomials. A thorough mathematical analysis shows that writing $f_n(E) = f_0(E) P_n(E)$ makes $\{P_n(E)\}$ a set of orthogonal polynomials satisfying (18) with a seed value of $P_0(E) = 1$. Thus, the wavefunction equation (12) become $\psi(E,x) = f_0(E) \sum_n P_n(E) |\phi_n(x)\rangle$, where $f_0(E)$ is an overall normalization factor whose square is proportional to the positive weight function associated with the orthogonal polynomial $\{P_n(E)\}$. With these, the wavefunction $\psi(E,x)$ is exactly determined and realized and the associated quantum system is well defined. Have known the orthogonal polynomials $\{P_n(E)\}$, how do we get the energy spectrum of the bound states, scattering phase shift of the continuum energy states, and so on? All these will be gotten from the asymptotics $(n \to \infty)$ of these polynomials which general takes the form

$$P_n^\mu(E) \approx n^{-\tau} A^\mu(E) \times \cos\left[ n^\xi \theta(E) + \delta^\mu(E) \right] \quad (19)$$

where $\tau$ and $\xi$ are real positive constants that depends on the particular polynomial. $A^\mu(E)$ is the scattering amplitude and $\delta^\mu(E)$ is the scattering phase shift. Now, the bound states (finite or infinite) if they exist, occurs at energies $\{E_m\}$ that makes the scattering amplitude vanish, $A^\mu(E_m) = 0$. How do we get the asymptotics of these orthogonal polynomials? There are various methods that could be used such as: Darboux method [9], Recurrence Relation [10] and so on. However, in this paper all relevant orthogonal polynomials asymptotics will be provided.

Table 1 contains well-known exactly solvable potential functions in conventional quantum mechanics. The quantum system described by each of these potential functions has explicit analytical expression for their energy spectrum, wavefunction, scattering phase shift and so on. In conventional quantum mechanics, these properties are obtained by solving the Schrodinger equation (11), after inserting the potential function $V(x)$, using separation of variables of the partial differential equation to distinct ordinary differential equations. Another method of finding the analytic solution of Schrodinger equation with solvable potential function is by reducing it to a given hypergeometric equation [11] whose solutions can be mapped, most of the time, to a classical orthogonal polynomials. Others method includes: supersymmetric quantum mechanics [12], potential algebra [13], path integration [14], and point canonical transformations [15].

The point canonical transformations provide an excellent mapping/grouping of the exact solvable potential functions in two classes. Potential functions in class I are the Coulomb and Morse which can be mapped into three-dimensional harmonic oscillator and their eigenfunctions correspond to the confluent hypergeometric functions which can be written as Laguerre polynomials. Also potential functions in class II are the Rosen – Morse (hyperbolic and trigonometric), Eckart, Poschl – Teller (I and II) which can be mapped into the generalized Scarf potential and their eigenfunctions correspond to the hypergeometric functions which can be written as associated Legendre functions.

The idea of PCT Implies that each potential function in a particular class can be mapped to the *generalized potential function (here we assumed that the energy spectrum and wavefunction are known)* of that class using a specific coordinate transformation through which we get the *energy spectrum and wavefunction of the former potential functions*.

Table 1: Some exactly Solvable Potential functions in Quantum mechanics with their associated properties. $H(x)$, $L_n^{(\alpha)}(x)$, and $P_n^{(\mu,\nu)}(x)$ are the Hermite, Laguerre and Jacobi polynomials. Also $\eta$, $\alpha$, $\beta$, $A$, $B$, $\tilde{A}$, $\tilde{B}$, $s$, $\lambda$, and $a$ are physical parameters. Based on conventional units used $\hbar = m = 1$ or $\hbar = 2m = 1$; there will be a change in form for each potential function and its associated properties.

| S/N | Name of Potential Function | Potential function $V(x)$ | Energy Spectrum $E_n(\eta)$ | Variable $y_\beta(x)$ | Wavefunction (Radial Part) $\psi_n(x)$ |
|---|---|---|---|---|---|
| 1 | Harmonic oscillator | $\frac{1}{2}m\bar{\omega}^2 r^2 + \frac{\tilde{l}(\tilde{l}+1)\hbar^2}{2mr^2} - \left(\tilde{l}+\frac{3}{2}\right)\hbar\bar{\omega}$  $0 \leq r < \infty$ | $2n\hbar\bar{\omega}$ | $y = m\bar{\omega}r^2$ | $\exp\left(-\frac{1}{2}y\right) y^{(\tilde{l}+1)/2} \times L_n^{\tilde{l}+\frac{1}{2}}(y)$ |
| 2 | Coulomb | $-\frac{e^2}{x} + \frac{l(l+1)\hbar^2}{2mx^2} + \frac{me^4}{2(l+1)^2 \hbar^2}$  $0 \leq x < \infty$ | $\frac{me^4}{2\hbar^2}\left(\frac{1}{(l+1)^2} - \frac{1}{(n+l+1)^2}\right)$ | $y = \frac{2me^2 x}{\hbar(n+l+1)}$ | $\exp\left(-\frac{1}{2}y\right) y^{l+1} \times L_n^{2l+1}(y)$ |
| 3 | Morse | $A^2 + B^2 e^{-2\alpha x} - 2B\left(A + \frac{\alpha\hbar}{2\sqrt{2m}}\right)e^{-\alpha x}$  $-\infty < x < \infty$ | $A^2 - \left(A - \frac{n\alpha\hbar}{\sqrt{2m}}\right)^2$ | $y = \frac{2\sqrt{2m}B}{\alpha} e^{-\alpha x}$ | $y^{(\sqrt{2m}A/\hbar\alpha - n)} \exp\left(-\frac{1}{2}y\right) \times L_n^{(2\sqrt{2m}A/\hbar\alpha - 2n)}(y)$ |
| 4 | Trigonometric Scarf potential | $-\tilde{A}^2 + \left(\tilde{A}^2 + \tilde{B}^2 - \frac{\tilde{A}\alpha\hbar}{\sqrt{2m}}\right)\cosec^2\alpha x$  $-\tilde{B}\left(2\tilde{A} - \frac{\alpha\hbar}{\sqrt{2m}}\right)\cot\alpha x \cosec\alpha x$  $0 \leq \alpha x \leq \pi$ | $\left(\tilde{A} - \frac{n\alpha\hbar}{\sqrt{2m}}\right)^2 - \tilde{A}^2$ | $y = \cos\alpha x$  $s = \frac{\sqrt{2m}}{\hbar}\frac{\tilde{A}}{\alpha}$  $\lambda = \frac{\sqrt{2m}}{\hbar}\frac{\tilde{B}}{\alpha}$ | $(1-y)^{(s-\lambda)/2} \times (1+y)^{(s+\lambda)/2} \times P_n^{\left(s-\lambda-\frac{1}{2}, s+\lambda-\frac{1}{2}\right)}(y)$ |
| 5 | Hyperbolic Scarf potential | $A^2 + \left(B^2 - A^2 - \frac{A\alpha\hbar}{\sqrt{2m}}\right)\sech^2\alpha x$  $+B\left(2A + \frac{\alpha\hbar}{\sqrt{2m}}\right)\sech\alpha x \tanh\alpha x$  $-\infty < x < \infty$ | $A^2 - \left(A - \frac{n\alpha\hbar}{\sqrt{2m}}\right)^2$ | $y = \sinh\alpha x$  $s = \frac{\sqrt{2m}}{\hbar}\frac{A}{\alpha}$  $\lambda = \frac{\sqrt{2m}}{\hbar}\frac{B}{\alpha}$ | $(1+y^2)^{-s/2} \times \exp(-\lambda\tan^{-1}y) \times P_n^{\left(-s-i\lambda-\frac{1}{2}, s+i\lambda-\frac{1}{2}\right)}(iy)$ |
| 6 | Rosen- Morse I | $-A^2 + \frac{B^2}{A^2} + 2B\tan\alpha x$  $+A\left(A - \frac{\alpha\hbar}{\sqrt{2m}}\right)\sec^2\alpha x$  $-\infty < x < \infty$ | $\left(A + \frac{n\alpha\hbar}{\sqrt{2m}}\right)^2 - A^2$  $+\frac{B^2}{A^2} - \frac{B^2}{\left(A + \frac{n\alpha\hbar}{\sqrt{2m}}\right)^2}$ | $y = \tan\alpha x$  $s = \frac{\sqrt{2m}}{\hbar}\frac{A}{\alpha}$  $\lambda = \frac{\sqrt{2m}}{\hbar}\frac{B}{\alpha}$  $a = \frac{\sqrt{2m}\lambda}{\hbar\alpha(s+n)}$ | $(1+y^2)^{-(s+n)/2} \times \exp(-a\tan^{-1}y) \times P_n^{(-s-n+ia, s-n-ia)}(-iy)$ |
| 7 | Rosen - Morse II | $A^2 + \frac{B^2}{A^2} + 2B\tanh\alpha x$  $-A\left(A + \frac{\alpha\hbar}{\sqrt{2m}}\right)\sech^2\alpha x$  $-\infty < x < \infty$ | $A^2 - \left(A - \frac{n\alpha\hbar}{\sqrt{2m}}\right)^2$  $+\frac{B^2}{A^2} - \frac{B^2}{\left(A - \frac{n\alpha\hbar}{\sqrt{2m}}\right)^2}$ | $y = \tanh\alpha x$  $s = \frac{\sqrt{2mA}}{\hbar\alpha}$  $\lambda = \frac{\sqrt{2mB}}{\hbar\alpha}$  $a = \frac{\sqrt{2m}\lambda}{\hbar\alpha(s-n)}$ | $(1-y)^{(s-n+a)/2} \times (1+y)^{(s-n-a)/2} \times P_n^{(s-n+a, s-n-a)}(y)$ |
| 8 | Rosen – Morse II | $A^2 + \left(B^2 + A^2 + \frac{A\alpha\hbar}{\sqrt{2m}}\right)\cosech^2\alpha x$  $-B\left(2A + \frac{\alpha\hbar}{\sqrt{2m}}\right)^2 \coth\alpha x \times \cosech\alpha x$  $0 \leq x < \infty$ | $A^2 - \left(A - \frac{n\alpha\hbar}{\sqrt{2m}}\right)^2$ | $y = \cosh\alpha x$  $s = \frac{\sqrt{2mA}}{\hbar\alpha}$  $\lambda = \frac{\sqrt{2mB}}{\hbar\alpha}$ | $(y-1)^{(\lambda-s)/2} \times (1+y)^{-(\lambda+s)/2} \times P_n^{\left(\lambda-s-\frac{1}{2}, -\lambda-s-\frac{1}{2}\right)}(y)$ |

| 9 | Eckart | $A^2 + \frac{B^2}{A^2} - 2B\coth\alpha x$ $+ A\left(A - \frac{\alpha\hbar}{\sqrt{2m}}\right)co\operatorname{sech}^2\alpha x$ $0 \leq x < \infty$ | $A^2 - \left(A + \frac{n\alpha\hbar}{\sqrt{2m}}\right)^2$ $+ \frac{B^2}{A^2} - \frac{B^2}{\left(A + \frac{n\alpha\hbar}{\sqrt{2m}}\right)^2}$ | $y = \coth\alpha x$ $s = \frac{\sqrt{2m}A}{\hbar\alpha}$ $\lambda = \frac{\sqrt{2m}B}{\hbar\alpha}$ $a = \frac{\sqrt{2m}\lambda}{\hbar(s-n)}$ | $(y-1)^{-(s+n-a)/2} \times (1+y)^{-(s+n+a)/2} \times P_n^{(-s+a-n,-s-a-n)}(y)$ |
|---|---|---|---|---|---|
| 10 | Poschl-Teller I | $-(A+B)^2 + A\left(A - \frac{\alpha\hbar}{\sqrt{2m}}\right)\sec^2\alpha x$ $+ B\left(B - \frac{\alpha\hbar}{\sqrt{2m}}\right)\cos ec^2\alpha x$ $0 \leq \alpha x \leq \frac{\pi}{2}$ | $\left(A + B + \frac{2n\alpha\hbar}{\sqrt{2m}}\right)^2 - (A+B)^2$ | $y = 1 - 2\sin^2\alpha x$ $s = \frac{A}{\alpha}$ $\lambda = \frac{B}{\alpha}$ | $(1-y)^{\lambda/2} \times (1+y)^{s/2} \times P_n^{\left(\lambda - \frac{1}{2}, s - \frac{1}{2}\right)}(y)$ |
| 11 | Poschl-Teller II | $(A-B)^2 - A\left(A + \frac{\alpha\hbar}{\sqrt{2m}}\right)\operatorname{sech}^2\alpha x$ $+ B\left(B - \frac{\alpha\hbar}{\sqrt{2m}}\right)\cos ech^2\alpha x$ $0 \leq x < \infty$ | $(A-B)^2 - \left(A + B + \frac{2n\alpha\hbar}{\sqrt{2m}}\right)^2$ | $y = 1 + 2\sinh^2\alpha x$ $s = \frac{A}{\alpha}$ $\lambda = \frac{B}{\alpha}$ | $(1-y)^{\lambda/2} \times (1+y)^{-s/2} \times P_n^{\left(\lambda - \frac{1}{2}, -s - \frac{1}{2}\right)}(y)$ |

So the associated properties (energy spectrum and wavefunction) of potential functions 2 and 3 can be gotten from the generalized potential function 1 using a specific transformation whereas the associated properties of the potential functions (5-11) can be gotten from the generalized potential function 4.

However, these potential functions are also solvable using the TRA without inserting them in the Schrodinger equation. This is possible because in the TRA, every exactly solvable potential function has a specific space transformation and basis element that will recover them with their physical properties: energy spectrum, wavefunction, scattering phase shift and so on. So here comes the question; how do we know the specific space transformation compatible that will result into a specific potential function? The answer is based on experience.

Later in section 4, we will give a table that will show each space transformation that corresponds to each potential function. As earlier said, we will assumed that readers of this paper are familiar with conventional quantum mechanics, and have the knowledge of solving the Schrödinger equation when the potential function is inserted particularly for exactly solvable potential functions

### 3. Basis element and Space transformation

Considering the domain of each solvable potential functions (with nonconventional ones) with respect to the Schrodinger equation; we basically use two basis elements: Jacobi basis and Laguerre basis. Each of these basis elements (with the right space transformation) takes (11) to new coordinate space $y(x)$ suitable for each. Which are either finite $[-1,+1]$ for the Jacobi basis or semi-finite $[0,\infty]$ for Laguerre basis.

**Jacobi basis**: this basis element is given as

$$|\phi_n(x)\rangle = A_n (1+y)^\alpha (1-y)^\beta P_n^{(\mu,\nu)}(y) \tag{20}$$

where $A_n = \sqrt{\frac{\lambda(2n+\mu+\nu+1)\Gamma(n+1)\Gamma(n+\mu+\nu+1)}{2^{\mu+\nu+1}\Gamma(n+\nu+1)\Gamma(n+\mu+1)}}$, in choosing $A_n$ sometime insight is required based on the derivatives of the configuration space transformation coordinate. $\alpha, \beta \geq 0$, $\mu, \nu > -1$. $P_n^{(\mu,\nu)}(y)$ is the Jacobi Polynomial that satisfies the following properties (which will use later):

$$P_n^{(\mu,\nu)}(y) = \frac{\Gamma(n+\mu+1)}{\Gamma(n+1)\Gamma(\mu+1)} {}_2F_1\left(-n, n+\mu+\nu+1; \mu+1; \frac{1-y}{2}\right) = (-1)^n P_n^{(\mu,\nu)}(-y) \tag{21}$$

$$\left\{(1-y^2)\frac{d^2}{dy^2} - [(\mu+\nu+2)y+\mu-\nu]\frac{d}{dy} + n(n+\mu+\nu+1)\right\}P_n^{(\mu,\nu)}(y) = 0 \tag{22}$$

$$(1-y^2)\frac{d}{dy}P_n^{(\mu,\nu)}(y) = -n\left(y+\frac{\nu-\mu}{2n+\mu+\nu}\right)P_n^{(\mu,\nu)}(y) + 2\frac{(n+\mu)(n+\nu)}{2n+\mu+\nu}P_{n-1}^{(\mu,\nu)}(y) \tag{23}$$

$$\int_{-1}^{+1}(1-y)^\mu(1+y)^\nu P_n^{(\mu,\nu)}(y)P_m^{(\mu,\nu)}(y)\,dy = \frac{2^{\mu+\nu+1}}{2n+\mu+\nu+1}\frac{\Gamma(n+\mu+1)\Gamma(n+\nu+1)}{\Gamma(n+1)\Gamma(n+\mu+\nu+1)}\delta_{nm} \tag{24}$$

$$\left(\frac{1\pm y}{2}\right)P_n^{(\mu,\nu)}(y) = \frac{2n(n+\mu+\nu+1)+(\mu+\nu)\left(\frac{\mu+\nu}{2}\pm\frac{\nu-\mu}{2}+1\right)}{(2n+\mu+\nu)(2n+\mu+\nu+2)}P_n^{(\mu,\nu)}(y)$$
$$\pm\frac{(n+\mu)(n+\nu)}{(2n+\mu+\nu)(2n+\mu+\nu+1)}P_{n-1}^{(\mu,\nu)}(y) \pm \frac{(n+1)(n+\mu+\nu+1)}{(2n+\mu+\nu+1)(2n+\mu+\nu+2)}P_{n+1}^{(\mu,\nu)}(y) \tag{25}$$

$$yP_n^{(\mu,\nu)}(y) = \frac{\nu^2-\mu^2}{(2n+\mu+\nu)(2n+\mu+\nu+2)}P_n^{(\mu,\nu)}(y) + \frac{2(n+\mu)(n+\nu)}{(2n+\mu+\nu)(2n+\mu+\nu+1)}P_{n-1}^{(\mu,\nu)}(y)$$
$$+\frac{2(n+1)(n+\mu+\nu+1)}{(2n+\mu+\nu+1)(2n+\mu+\nu+2)}P_{n+1}^{(\mu,\nu)}(y) \tag{26}$$

A look at the orthogonality relation of the Jacobi polynomial shows that the polynomial is defined on the interval $[-1,+1]$. So to use the Jacobi basis, we must transform the (11) to new space coordinate. Suppose we intend to solve (11) with a potential function $V(x)$ defined on $[0,\infty]$; we choose $y = 1 - 2e^{-\lambda x}$, for instance, which will change the space configuration of (11) from $[0,\infty]$ to $[-1,+1]$. In general for any space transformation, the Schrodinger equation (11) in conventional units of $\hbar = m = 1$ becomes.

$$(H-E)|\psi\rangle = -\frac{1}{2}\left[(y')^2\frac{d^2}{dy^2} + y''\frac{d}{dy} - 2V(y) + 2E\right]|\psi\rangle = 0 \tag{27}$$

where we have used $\frac{d}{dx} = y'\frac{d}{dy}$ and $\frac{d^2}{dx^2} = (y')^2\frac{d^2}{dy^2} + y''\frac{d}{dy}$. Now using $y = 1 - 2e^{-\lambda x}$, (27) becomes

$$(H-E)|\psi\rangle = -\frac{1}{2}\left[4\lambda^2 e^{-2\lambda x}\frac{d^2}{dy^2} - 2\lambda^2 e^{-\lambda x}\frac{d}{dy} - 2V(y) + 2E\right]|\psi\rangle = 0$$
$$= -\frac{1}{2}\left[\lambda^2(1-y)^2\frac{d^2}{dy^2} - \lambda^2(1-y)\frac{d}{dy} - 2V(y) + 2E\right]|\psi\rangle = 0 \tag{28}$$

since $e^{-\lambda x} = \frac{1-y}{2}$ and $e^{-2\lambda x} = \frac{(1-y)^2}{4}$. Now we use (12) or invariably say we operate the wave operator $J = (H-E)$ on the basis element (left and right hand side), and then we have

$$-\frac{2}{\lambda^2}\sum_{n=0}^\infty f_n(\varepsilon)\langle\phi_m(\varepsilon)|(H-E)|\phi_n(y)\rangle = \sum_{n=0}^\infty f_n(\varepsilon)\langle\phi_m(\varepsilon)|\left(\left[(1-y)^2\frac{d^2}{dy^2} - (1-y)\frac{d}{dy} - \frac{2V(y)}{\lambda^2} + \frac{2E}{\lambda^2}\right]|\phi_m(y)\rangle\right) = 0 \tag{29}$$

from this it can be seen that the wave operator matrix elements is

$$J_{nm} = -\frac{2}{\lambda^2}\langle\phi_m(y)|(H-E)|\phi_n(y)\rangle = \langle\phi_m(\varepsilon)|\left(\left[\lambda^2(1-y)^2\frac{d^2}{dy^2} - \lambda^2(1-y)\frac{d}{dy} - \frac{2V(y)}{\lambda^2} + \frac{2E}{\lambda^2}\right]|\phi_m(y)\rangle\right) \tag{30}$$

The use of the differential, recursion, and orthogonality properties of the Jacobi polynomial in this equation will result in $J_{n,n}f_n + J_{n,n+1}f_{n+1} + J_{n,n-1}f_{n-1} = 0$. Comparism with well-known recursion relation of classical orthogonal polynomial will give the expansion coefficient $f_n(\varepsilon)$ of the wavefunction. Evaluation of (30) and the requirement of it to be tridiagonal and symmetric matrix is the major pivot of the TRA. Also from (29), without inserting the transformed potential function $V(y)$; and using the Jacobi basis, we can proceed to get wavefunction and other associated properties of the quantum systems. Interestingly, we will recover $V(y)$ (hence $V(x)$) and others well - known (unknown) solvable potential functions that are defined on $[0,\infty]$. This is awesome power of the TRA. So, from (29) we proceed until we achieved (14-18) and we finally use (12). Detailed steps will be shown later. Most time when using the Jacobi basis element (or Laguerre basis), we usually write (27) in a convenient form so as to be able to use the properties (21-26) of the Jacobi orthogonal polynomials. For instance using our previous space transformation $y = 1 - 2e^{-\lambda x}$, (27) could be written as

$$\sum_{n=0}^{\infty} f_n(\varepsilon)\langle\phi_m(y)|(H-E)|\phi_n(y)\rangle = \sum_{n=0}^{\infty} f_n(\varepsilon)\langle\phi_m(y)|\left\{-\frac{1}{2}\frac{(y')^2}{(1-y^2)}\left((1-y^2)\frac{d^2}{dy^2} + \frac{y''}{(y')^2}(1-y^2)\frac{d}{dy} + 2\frac{(1-y^2)}{(y')^2}[E-V]\right)|\phi_n(y)\rangle\right\} = 0 \quad (31)$$

and we further get

$$J_{nm} = \langle\phi_m(y)|(H-E)|\phi(y)\rangle = \langle\phi_m(y)|\left\{-\frac{\lambda^2}{2}\frac{(1-y)}{(1+y)}\left((1-y^2)\frac{d^2}{dy^2} - (1+y)\frac{d}{dy} + \frac{2(1+y)}{\lambda^2(1-y)}[E-V]\right)|\phi_n(y)\rangle\right\} = 0 \quad (32)$$

where we have used $y' = \lambda(1-y)$ and $\frac{y''}{(y')^2} = \frac{-1}{(1-y)}$. However, this does not affect the correctness of (27) and (29). In fact equation (27) will always be our starting point later. For clarity sake in our later computation we will only be writing $J|\phi_n(y)\rangle = (H-E)|\phi_n(y)\rangle$ and generalized thereafter. This does not affect our work. From (31), there will be a need for the second (first) differential operator of the Jacobi basis element. These are given as

$$\frac{d}{dy}\phi_n(y) = A_n(1-y)^\alpha(1+y)^\beta\left(\frac{d}{dy} + \frac{\beta}{1+y} - \frac{\alpha}{1-y}\right)P_n^{(u,v)}(y) \quad (33)$$

and

$$\frac{d^2}{dy^2}\phi_n(y) = A_n(1-y)^\alpha(1+y)^\beta\left(\frac{d^2}{dy^2} + 2\left[\frac{\beta}{1+y} - \frac{\alpha}{1-y}\right]\frac{d}{dy} + \frac{\beta(\beta-1)}{(1+y)^2} + \frac{\alpha(\alpha-1)}{(1-y)^2} - \frac{2\alpha\beta}{1-y^2}\right)P_n^{(u,v)}(y) \quad (34)$$

In order to use the orthogonality property (24) of the Jacobi orthogonal polynomial, the integration measure is always specified base on the space transformation used. Like the above sample we have $\int_0^\infty dx.... = \int_{-1}^{+1}\frac{dy}{\lambda(1-y)}....$

As another example, a potential function defined on $[-\infty,\infty]$ can be transformed into a new space coordinate using $y = \tanh(\lambda x)$ (this compatible with the Jacobi basis element), then (27) becomes

$$(H-E)|\psi\rangle = -\frac{\lambda^2(1-y^2)}{2}\left[(1-y^2)\frac{d^2}{dy^2} - 2y\frac{d}{dy} - \frac{2V(y)}{\lambda^2(1-y^2)} + \frac{2E}{\lambda^2(1-y^2)}\right]|\psi\rangle = 0 \quad (35)$$

since $y' = \lambda(1-y^2)$ and $y'' = -2\lambda^2 y(1-y^2)$. Similarly, we have

$$-\frac{2}{\lambda^2}\sum_{n=0}^{\infty} f_n(\varepsilon)\langle\phi_m(y)|(H-E)|\phi_n(y)\rangle = \sum_{n=0}^{\infty} f_n(\varepsilon)\langle\phi_m(y)|\left\{(1-y^2)\left[(1-y^2)\frac{d^2}{dy^2} - 2y\frac{d}{dy} - \frac{2V(y)}{\lambda^2(1-y^2)} + \frac{2E}{\lambda^2(1-y^2)}\right]\right\}|\phi_n(y)\rangle = 0 \quad (36)$$

And the wave operator matrix element is
;

$$J_{nm} = -\frac{2}{\lambda^2}\langle\phi_m(y)|(H-E)|\phi_n(y)\rangle = \langle\phi_m(y)|\left\{(1-y^2)\left[(1-y^2)\frac{d^2}{dy^2} - 2y\frac{d}{dy} - \frac{2V(y)}{\lambda^2(1-y^2)} + \frac{2E}{\lambda^2(1-y^2)}\right]\right\}|\phi_n(y)\rangle = 0 \quad (37)$$

with integration measure $\int_{-\infty}^{\infty} dx.... = \int_{-1}^{+1}\frac{dy}{\lambda(1-y^2)}....$

**Laguerre Basis:** this basis element is given as

$$|\phi_n(x)\rangle = A_n y^\alpha e^{-\beta y} L_n^v(y) \quad (38)$$

~ 9 ~

where $\alpha$ and $v$ are real parameters with $v > -1$ and $\alpha \geq 0$ to ensure convergence of the Laguerre polynomial $L_n^v(y)$ and compatibility with boundary conditions when the new variable $y$ span semi-infinite interval $[0,\infty]$. Also $A_n = \sqrt{\lambda \Gamma(n+1)/\Gamma(n+v+1)}$, in choosing $A_n$ sometime insight is required based on the derivatives of the configuration space transformation coordinate. The Laguerre polynomial $L_n^v(y)$ satisfies the following properties

$$yL_n^v(y) = (2n+v+1)L_n^v(y) - (n+v)L_{n-1}^v(y) - (n+1)L_{n+1}^v(y) \tag{39}$$

$$L_n^v(y) = \frac{\Gamma(n+v+1)}{\Gamma(n+1)\Gamma(v+1)} {}_1F_1(-n; v+1; y) \tag{40}$$

$$\left[ y \frac{d^2}{dy^2} + (v+1-y) \frac{d}{dy} + n \right] L_n^v(y) = 0 \tag{41}$$

$$y \frac{d}{dy} L_n^v(y) = n L_n^v(y) - (n+v) L_{n-1}^v(y) \tag{42}$$

$$\int_0^\infty y^v e^{-y} L_n^v(y) L_m^v(y) dy = \frac{\Gamma(n+v+1)}{\Gamma(n+1)} \delta_{nm} \tag{43}$$

Now to use the Laguerre basis, we must transform the (11) to new space coordinate. Suppose we intend to solve (11) with a potential function $V(x)$ defined on $[-\infty,\infty]$; we can choose $y = \mu e^{-\lambda x}$, for instance, which will change the space configuration of (11) from $[-\infty,\infty]$ to $[0,\infty]$ which is compatible with the Laguerre basis. Then (27) becomes

$$(H-E)|\psi\rangle = -\frac{\lambda^2}{2}\left[ y^2 \frac{d^2}{dy^2} + y \frac{d}{dy} - \frac{2V(y)}{\lambda^2} + \frac{2E}{\lambda^2} \right]|\psi\rangle = 0 \tag{44}$$

since $y' = -\lambda y$ and $y'' = \lambda^2 y$. Moving further we get

$$-\frac{2}{\lambda^2} \sum_{n=0}^\infty f_n(\varepsilon) \langle \phi_m(y)|(H-E)|\phi_n(y)\rangle = \sum_{n=0}^\infty f_n(\varepsilon) \langle \phi_m(y)| \left[ y^2 \frac{d^2}{dy^2} + y \frac{d}{dy} - \frac{2V(y)}{\lambda^2} + \frac{2E}{\lambda^2} \right] |\phi_n(y)\rangle = 0 \tag{45}$$

and the wave operator matrix element is

$$J_{nm} = -\frac{2}{\lambda^2} \langle \phi_m(y)|(H-E)|\phi_n(y)\rangle = \langle \phi_m(y)| \left[ y^2 \frac{d^2}{dy^2} + y \frac{d}{dy} - \frac{2V(y)}{\lambda^2} + \frac{2E}{\lambda^2} \right] |\phi_n(y)\rangle = 0 \tag{46}$$

with integral measure $\int_{-\infty}^\infty dx .... = \int_0^\infty \frac{dy}{\lambda y} .....$. Also the first (second) order differential operator of the Laguerre basis are

$$\frac{d}{dy}|\phi_n(y)\rangle = A_n y^\alpha e^{-\beta y} \left[ \frac{d}{dy} + \frac{\alpha}{y} - \beta \right] L_n^v(y) \tag{47}$$

and

$$\frac{d^2}{dy^2}|\phi_n(y)\rangle = A_n y^\alpha e^{-\beta y} \left[ \frac{d^2}{dy^2} + \left( \frac{2\alpha}{y} - 2\beta \right) \frac{d}{dy} - \frac{\alpha}{y^2} + \left( \frac{\alpha}{y} - \beta \right)^2 \right] L_n^v(y) \tag{50}$$

Similar inference like the Jacobi basis can be made for the Laguerre basis. In conclusion, all solvable potential functions in quantum mechanics can be recovered (with new ones) in the TRA. Each of the potential functions has specific space transformation (compatible with Jacobi basis or Laguerre basis) that can be used in transforming the Schrödinger equation and we proceed from either (31) or (45) without inserting the transformed potential function $V(y)$ which we will later recover with associated properties. So as students of quantum mechanics require to solve the Schrodinger equation with a solvable potential function, in order to use the TRA, you are only require to know the right space transformation that will produce this potential function and its associated properties. As an exercise, we advise readers to look at some space transformation that are compatible with either the Jacobi basis or Laguerre basis considering the potential functions in table 1.

## 4. Classical orthogonal polynomial

A lot of literatures exist on orthogonal polynomials. For quick reference, we advise readers to look at [16] where various types of orthogonal polynomials and their properties are well defined. In [16] the normalized versions of these classical orthogonal polynomials are not given. See appendix A for normalized form of all orthogonal polynomials. From our experience in TRA, using the Laguerre bases the matrix wave equation lead to three-term recursion relations for the expansion coefficients of the continuum wavefunction. Most time, they are the recursion relations of either the Meixner- Pollaczek or the Continuous dual Hahn polynomials (are used as expansion coefficients for the bound states wavefunction). ***The solvable potential functions (quantum systems) when we will use this base are Coulomb, 3D isotropic (or 1D harmonic) oscillator, and the 1D Morse potential***. Now we give the properties of these polynomials and make reference to them when illustrating examples. The normalized version of the Meixner-Pollaczek polynomial is [Sec 1.7 of Ref 16]

$$P_n^\mu(z,\theta) = \sqrt{\frac{(2\mu)_n}{n!}} e^{in\theta} \, {}_2F_1\left(\begin{matrix} -n, \mu+iz \\ 2\mu \end{matrix} \middle| 1-e^{-2i\theta}\right) \tag{51}$$

where $(a)_n = a(a+1)(a+2)...(a+n-1) = \frac{\Gamma(n+a)}{\Gamma(a)}$, $z$ is the whole real line, $\mu > 0$ and $0 < \theta < \pi$. The orthogonality relation for this polynomial is

$$\int_{-\infty}^{\infty} \rho^\mu(z,\theta) P_n^\mu(z,\theta) P_m^\mu(z,\theta) dz = \delta_{nm} \tag{52}$$

where $\rho^\mu(z,\theta) = \frac{1}{2\pi\Gamma(2\mu)}(2\sin\theta)^{2\mu} e^{(2\theta-\pi)z} |\Gamma(\mu+iz)|^2$ is the normalized weight function. The three term recursive relation for this polynomial

$$(z\sin\theta) P_n^\mu(z,\theta) = -\left[(n+\mu)\cos\theta\right] P_n^\mu(z,\theta)$$
$$+ \frac{1}{2}\sqrt{n(n+2\mu-1)} P_{n-1}^\mu(z,\theta) + \frac{1}{2}\sqrt{(n+1)(n+2\mu)} P_{n+1}^\mu(z,\theta) \tag{53}$$

In [17], it was shown that the asymptotics of this polynomial is

$$P_n^\mu(z,\theta) \approx \frac{2n^{-1/2} e^{\left(\frac{\pi}{2}-\theta\right)z}}{(2\sin\theta)^\mu \Gamma(\mu+iz)} \cos\left[n\theta + \arg\Gamma(\mu+iz) - \mu\frac{\pi}{2} - z\ln(2n\sin\theta)\right] \tag{54}$$

Since $\ln n \approx o(n)$ as $n \to \infty$, we can neglect $z\ln(2n\sin\theta)$ relative to $n\theta$. The scattering amplitude is

$$A^\mu(\varepsilon) = \frac{2e^{\left(\frac{\pi}{2}-\theta\right)z}}{(2\sin\theta)^\mu \Gamma(\mu+iz)} \tag{55}$$

and phase shift is

$$\delta^\mu(\varepsilon) = \arg\Gamma(\mu+iz) - \mu\frac{\pi}{2} \tag{56}$$

The discrete infinite spectrum occurs if $\mu + iz = -m$, where $m = 0,1,2,...,N$. The spectrum formula associated with this polynomial is $z_m^2 = -(m+\mu)^2$. This spectrum is infinite for positive $\mu$, whereas it is finite for negative $\mu$ with $m = 0,1,...,N$ and $N$ is the largest integer less than or equal to $-\mu$. The generating function is

$$\sum_{n=0}^\infty \tilde{P}_n^\mu(z,\theta) t^n = (1-te^{i\theta})^{-\mu+iz} (1-te^{-i\theta})^{-\mu-iz} \tag{57}$$

where $\tilde{P}_n^\mu(z;\theta) = \sqrt{\frac{\Gamma(n+2\mu)}{\Gamma(2\mu)\Gamma(n+1)}} P_n^\mu(z;\theta)$. The discrete version of Meixner Pollaczek polynomial is obtained by substituting $z = i(m+\mu)$ and $\theta \to i\theta$ in (51) to get

$$M_n^\mu(m;\beta) = \sqrt{\frac{(2\mu)_n}{n!}} \beta^{n/2} {}_2F_1\left(\begin{matrix}-n,-m\\2\mu\end{matrix}\bigg|1-\beta^{-1}\right) \tag{58}$$

where $\beta = e^{-2\theta}$ with $\theta > 0$ making $0 < \beta < 1$. With this substitution and denoting $2\cosh\theta = \frac{1}{\sqrt{\beta}} + \sqrt{\beta}$ and $2\sinh\theta = \frac{1}{\sqrt{\beta}} - \sqrt{\beta}$, then the recursive relation (53) becomes

$$(\beta-1)mM_n^\mu(m;\beta) = -\left[n(1+\beta)+2\mu\beta\right]M_n^\mu(m;\beta) \tag{59}$$
$$+\sqrt{n(n+2\mu-1)\beta}M_{n-1}^\mu(m;\beta) + \sqrt{(n+1)(n+2\mu)\beta}M_{n+1}^\mu(m;\beta)$$

for the discrete Meixner Pollaczek polynomial with normalized discrete weight function $\omega_m^\mu(\beta) = (1-\beta)^{2\mu}\frac{\Gamma(m+2\mu)}{\Gamma(2\mu)\Gamma(m+1)}$, and orthogonality $\sum_{m=0}^{\infty}\omega_m^\mu(\beta)M_n^\mu(m;\beta)M_\alpha^\mu(m;\beta) = \delta_{n\alpha}$. Now suppose we take $2\mu = -N$, where $N$ is a non-negative integer and $\beta = \frac{-\gamma}{1-\gamma}$, then (51) will become the normalized discrete **Krawtchouk polynomial(finite spectrum)** defined as

$$K_n^\mu(m;\gamma) = \sqrt{\frac{N!}{n!(N-n)!}} \left(\frac{\gamma}{1-\gamma}\right)^{n/2} {}_2F_1\left(\begin{matrix}-n,-m\\-N\end{matrix}\bigg|\gamma^{-1}\right) \tag{60}$$

where $\gamma^{-1} = 1 - \beta^{-1}$ with $0 < \gamma < 1$ and $n, m = 0, 1, ..., N$. Note that we have used $(-N)_n = \frac{\Gamma(n-N)}{\Gamma(-N)} = (-1)^n\frac{\Gamma(N+1)}{\Gamma(N-n+1)}$ in getting (58) from (16). The recursive relation this polynomial will be

$$mK_n^\mu(m;\beta) = \left[N\gamma + n(1-2\gamma)\right]K_n^\mu(m;\gamma) \tag{61}$$
$$-\sqrt{n(N-n+1)\gamma(1-\gamma)}K_{n-1}^\mu(m;\gamma) + \sqrt{(n+1)(N-n)\gamma(1-\gamma)}K_{n+1}^\mu(m;\gamma)$$

with discrete normalized weight function $\omega_m^N(\gamma) = (1-\gamma)^{N-m}\frac{\Gamma(N+1)\gamma^m}{\Gamma(N-m+1)\Gamma(m+1)}$, and orthogonality $\sum_{m=0}^{N}\omega_m^N(\gamma)K_n^N(m;\gamma)K_\alpha^N(m;\gamma) = \delta_{n\alpha}$. The normalized version of the continuous dual Hahn polynomial is (Sec. 1.3 of Ref 38),

$$S_n^\mu(z^2;\alpha,\beta) = \sqrt{\frac{(\mu+\alpha)_n(\mu+\beta)_n}{n!(\alpha+\beta)_n}} {}_3F_2\left(\begin{matrix}-n,\mu+iz,\mu-iz\\\mu+\alpha,\mu+\beta\end{matrix}\bigg|1\right) \tag{62}$$

where ${}_3F_2\left(\begin{matrix}a,b,c\\d,e\end{matrix}\bigg|z\right) = \sum_{n=0}^{\infty}\frac{(a)_n(b)_n(c)_n}{(d)_n(e)_n}\frac{z^n}{n!}$ is the generalized hypergeometric function and $(c)_n = c(c+1)(c+2)...(c+n-1) = \frac{\Gamma(n+c)}{\Gamma(c)}$. If the parameters $\{\mu,\alpha,\beta\}$ are positive except for a complex conjugates with positive real parts, then the spectrum of this polynomial is purely continuous. However, if $\mu < 0$ and $\mu+\alpha$, $\mu+\beta$ are positive or complex conjugates with positive real part, then the spectrum is a mix of a continuous part and a discrete part. The polynomial satisfies the following symmetric three term recursion relation:

$$z^2 S_n^\mu = [(n+\mu+\alpha)(n+\mu+\beta)+n(n+\alpha+\beta-1)-\mu^2]S_n^\mu$$
$$-\sqrt{n(n+\alpha+\beta-1)(n+\mu+\alpha-1)(n+\mu+\beta-1)}S_{n-1}^\mu$$
$$-\sqrt{(n+1)(n+\alpha+\beta)(n+\mu+\alpha)(n+\mu+\beta)}S_{n+1}^\mu \tag{63}$$

The corresponding normalized weight function is

$$\rho^\mu(z;\alpha,\beta) = \frac{1}{2\pi}\frac{\left|\Gamma(\mu+iz)\Gamma(\alpha+iz)\Gamma(\beta+iz)/\Gamma(2iz)\right|^2}{\Gamma(\mu+\alpha)\Gamma(\mu+\beta)\Gamma(\alpha+\beta)} \tag{64}$$

The asymptotics $(n \to \infty)$ of this polynomial is

$$S_n^\mu(z^2;\alpha^2,\beta) \approx \frac{2\sqrt{\Gamma(\mu+\alpha)\Gamma(\mu+\beta)\Gamma(\alpha+\beta)}|\Gamma(2iz)|}{|\Gamma(\alpha+iz)\Gamma(\beta+iz)\Gamma(\mu+iz)|\sqrt{n}} \times \tag{65}$$

$$\cos\{z\ln n + \arg[\Gamma(2iz)/\Gamma(\mu+iz)\Gamma(\alpha+iz)\Gamma(\beta+iz)]\}$$

since $\ln n \approx o(n^\xi)$ for any $\xi > 0$. From (65) the scattering phase shift is

$$\delta^\mu(z) = \arg[\Gamma(2iz)/\Gamma(\mu+iz)\Gamma(\alpha+iz)\Gamma(\beta+iz)] \tag{66}$$

The scattering amplitude in (65) vanishes if $\mu + iz = -n$ which will give the finite (discrete) spectrum formula $z^2 = -(n+\mu)^2$, where $n = 0,1,2,\ldots,N$ and $N$ is the largest integer less than or equal to $-\mu$. The generating function is

$$\sum_{n=0}^\infty \tilde{S}_n^\mu(z^2;\alpha,\beta)t^n = (1-t)^{-\mu+iz} {}_2F_1\left(\begin{array}{c}\alpha+iz,\beta+iz\\ \alpha+\beta\end{array}\bigg|t\right) \tag{67}$$

where $S_n^\mu(z^2;\alpha,\beta) = \frac{(\mu+\alpha)_n(\mu+\beta)_n}{n!(\alpha+\beta)} {}_3F_2\left(\begin{array}{c}-n,\mu+iz,\mu-iz\\ \mu+\alpha,\mu+\beta\end{array}\bigg|1\right)$. The discrete version of the polynomial is the dual Hahn polynomial, which we write as (Sec. 1.3 of Ref 38)

$$R_n^N(m;\alpha,\beta) = \sqrt{\frac{(\alpha+1)_n(\beta+1)_{N-n}}{n!(N-n)!}} {}_3F_2\left(\begin{array}{c}-n,-m,m+\alpha+\beta+1\\ \alpha+1,-N\end{array}\bigg|1\right) \tag{68}$$

where $n,m = 0,1,2,\ldots,N$ and either $\alpha,\beta > -1$ or $\alpha,\beta < -N$. The three term recursion relation is

$$\left(m+\frac{\alpha+\beta+1}{2}\right)^2 R_n^N(m;\alpha,\beta) = -\left\{\left(n+\frac{\alpha+1}{2}\right)^2 + \left(n-\frac{\beta+1}{2}\right)^2 - N(2n+\alpha+1) - \frac{1}{4}\left[(\alpha+\beta+1)^2+(\alpha+1)^2+(\beta+1)^2\right]\right\} R_n^N(m;\alpha,\beta)$$

$$+\sqrt{n(n+\alpha)(N-n+1)(N-n+\beta+1)}R_{n-1}^N(m;\alpha,\beta) + \sqrt{(n+1)(n+\alpha+1)(N-n)(N-n+\beta)}R_{n+1}^N(m;\alpha,\beta) \tag{69}$$

The normalized discrete weight function is

$$\rho^N(m;\alpha,\beta) = (N!)\frac{(2m+\alpha+\beta+1)(\alpha+1)_m(N-m+1)_m}{(m+\alpha+\beta+1)_{N+1}(\beta+1)_m m!} \tag{70}$$

In the Jacobi bases, the matrix wave equation leads to three –term recursion relations for the expansion coefficients of the continuum wavefunction. However, these recursion relations and their associated orthogonal polynomials are not treated in the mathematics literature and their analytic properties (weight function, generating, orthogonality, differential property, spectrum formula, asymptotics, etc) are yet to be derived. The special cases of these polynomials are already encountered in physics [39]. ***The solvable potential functions (quantum systems) when we use this bases are Poschl –Teller, Scarf, Eckart, Rosen – Morse, etc) as well as new potentials or generalization of the conventional potential functions***. Now, we define these polynomials by their three – term recursion relations and also obtain the spectrum formula for special cases of these polynomials using the exact (finite/infinite) energy spectrum of the associated exactly solvable physical problems that belong to the conventional class.

We denote the first polynomial as $\bar{H}_n^{(\mu,v)}(z^{-1};\alpha,\theta)$, which is a four parameter orthogonal polynomial. It satisfies a three term recursion relation written as

$$(\cos\theta)\bar{H}_n^{(\mu,v)}(z^{-1};\alpha,\theta) = \left\{z^{-1}\sin\theta\left[\left(n+\frac{\mu+v+1}{2}\right)^2+\alpha\right] + \frac{v^2-\mu^2}{(2n+\mu+v)(2n+\mu+v+2)}\right\}\bar{H}_n^{(\mu,v)}(z^{-1};\alpha,\theta)$$

$$+\frac{2(n+\mu)(n+v)}{(2n+\mu+v)(2n+\mu+v+1)}\bar{H}_{n-1}^{(\mu,v)}(z^{-1};\alpha,\theta) + \frac{2(n+1)(n+\mu+v+1)}{(2n+\mu+v+1)(2n+\mu+v+2)}\bar{H}_{n+1}^{(\mu,v)}(z^{-1};\alpha,\theta) \tag{71}$$

where $n = 1,2,\ldots$, $0 \leq \theta \leq \pi$ and $\bar{H}_0^{(\mu,v)}(z^{-1};\alpha,\theta) = 1$. It is a polynomial of order $n$ in $z^{-1}$ and in $\alpha$. This recursion relation becomes that of the Jacobi polynomial $P_n^{(\mu,v)}(\cos\theta)$ when $z \to \infty$. The polynomial of the first kind satisfies this recursion relation with $\bar{H}_0^{(\mu,v)}(z^{-1};\alpha,\theta) = 1$ and

~ 13 ~

$$\bar{H}_{n}^{(\mu,v)}\left(z^{-1};\alpha,\theta\right)=\frac{\mu-v}{2}+\frac{1}{2}(\mu+v+2)\left\{\cos\theta-z^{-1}\sin\theta\left[\frac{1}{4}(\mu+v+1)^{2}+\alpha\right]\right\} \quad (72)$$

which is obtained from (71) with $n=0$ and $\bar{H}_{-1}^{(\mu,v)}\left(z^{-1};\alpha,\theta\right)\equiv 0$. The polynomial of the second kind satisfies the same recursion relation (71) with $\bar{H}_{0}^{(\mu,v)}\left(z^{-1};\alpha,\theta\right)=1$ but $\bar{H}_{1}^{(\mu,v)}\left(z^{-1};\alpha,\theta\right)=c_{0}+c_{1}z^{-1}$, where the linearity coefficients $c_{0}$ and/or $c_{1}$ are different from those in (72). The orthonormal version of $\bar{H}_{n}^{(\mu,v)}\left(z^{-1};\alpha,\theta\right)$ is defined as

$$H_{n}^{(\mu,v)}\left(z^{-1};\alpha,\theta\right)=A_{n}\bar{H}_{n}^{(\mu,v)}\left(z^{-1};\alpha,\theta\right) \quad \text{where} \quad A_{n}=\sqrt{\frac{2n+\mu+v+1}{\mu+v+1}\frac{n!(\mu+v+1)_{n}}{(\mu+1)_{n}(v+1)_{n}}} \quad \text{and}$$

$(a)_{n}=a(a+1)(a+2)...(a+n-1)=\frac{\Gamma(n+a)}{\Gamma(a)}$. This polynomial has only continuous spectrum over the whole real $z$ line. The asymptotic behaviour $(n\to\infty)$ of $H_{n}^{(\mu,v)}\left(z^{-1};\alpha,\theta\right)$, numerically using (71), is sinusoidal and consistent with (19) when $\tau=\frac{1}{2}$ and $\xi=1$. The asymptotics of this polynomial is not known like (54) hence the physical features (phase shift and energy spectrum) of the corresponding quantum mechanical system could be obtained analytically. However, when we use the Jacobi base with a specific space transformation we often get the recursion relation (71) with the corresponding specific potential function and numerically the eigenvalues hence the plot of the wavefunction using (16) and (12). Also, this polynomial has two discrete versions – one with infinite discrete spectrum and the other with finite spectrum. Putting $\theta\to i\theta$ and $z\to iz_{k}$, where $k$ is an integer of either finite or infinite range, (71) becomes

$$(1+\beta)\bar{H}_{n}^{(\mu,v)}(k;\alpha,\beta)=\left(z_{k}^{-1}(1-\beta)\left[\left(n+\frac{\mu+v+1}{2}\right)^{2}+\alpha\right]+\frac{2(v^{2}-\mu^{2})\sqrt{\beta}}{(2n+\mu+v)(2n+\mu+v+2)}\right)\bar{H}_{n}^{(\mu,v)}(k;\alpha,\beta)$$

$$+\frac{4(n+\mu)(n+v)\sqrt{\beta}}{(2n+\mu+v)(2n+\mu+v+1)}\bar{H}_{n-1}^{(\mu,v)}(k;\alpha,\beta)+\frac{4(n+1)(n+\mu+v+1)\sqrt{\beta}}{(2n+\mu+v+1)(2n+\mu+v+2)}\bar{H}_{n+1}^{(\mu,v)}(k;\alpha,\beta) \quad (73)$$

where $\beta=e^{-2\theta}$ with $\theta>0$.

We designated the second as $\bar{G}_{n}^{(\mu,v)}\left(z^{2};\sigma\right)$ and it satisfies the three term recursion relation for $n=1,2,..$

$$z^{2}\bar{G}_{n}^{(\mu,v)}\left(z^{2};\sigma\right)=\left\{\left(\sigma+B_{n}^{2}\right)\left[\frac{2(n+\mu)(n+v)}{(2n+\mu+v)(2n+\mu+v+2)}+1\right]-\frac{2n(n+v)}{2n+\mu+v}-\frac{1}{2}(\mu+1)^{2}\right\}\bar{G}_{n}^{(\mu,v)}\left(z^{2};\sigma\right)$$

$$-\left(\sigma+B_{n-1}^{2}\right)\frac{2(n+\mu)(n+v)}{(2n+\mu+v)(2n+\mu+v+1)}\bar{G}_{n-1}^{(\mu,v)}\left(z^{2};\sigma\right)-\left(\sigma+B_{n}^{2}\right)\frac{2(n+1)(n+\mu+v+1)}{(2n+\mu+v+1)(2n+\mu+v+2)}\bar{G}_{n-1}^{(\mu,v)}\left(z^{2};\sigma\right) \quad (74)$$

where $B_{n}=\left(n+\frac{\mu+v}{2}+1\right)$ and $\bar{G}_{0}^{(\mu,v)}\left(z^{2};\sigma\right)=1$. It is a polynomial of order $n$ in $z^{2}$. The polynomial of the first kind satisfies this recursion relation with $\bar{G}_{0}^{(\mu,v)}\left(z^{2};\sigma\right)=1$ and

$$\bar{G}_{1}^{(\mu,v)}\left(z^{2};\sigma\right)=\mu+1-(\mu+v+2)\frac{z^{2}+\frac{1}{2}(\mu+1)^{2}}{2(\sigma+B_{0}^{2})}. \quad (75)$$

The second kind satisfies (73) with $\bar{G}_{0}^{(\mu,v)}\left(z^{2};\sigma\right)=1$ but with $\bar{G}_{1}^{(\mu,v)}\left(z^{2};\sigma\right)=c_{0}+c_{1}z^{2}$ and the linear coefficients $\{c_{0},c_{1}\}$ are different those (75). When $\sigma$ is positive this polynomial has a continuous spectrum on the positive $z^{2}$ line but if negative the spectrum is a mix of a continuous part on the positive $z^{2}$ line and a discrete part on the negative $z^{2}$ line numerically using (73) as $n$ becomes larger. The asymptotic behaviour $(n\to\infty)$ of $\bar{G}_{n}^{(\mu,v)}\left(z^{2};\sigma\right)$, numerically using (73), is sinusoidal and consistent with (19) when $\tau=\frac{1}{2}$ with $n^{\xi}\to\ln(z)$. That is, the asymptotic limit takes the form

$$\bar{G}_{n}^{(\mu,v)}\left(z^{2};\sigma\right)\approx\frac{1}{\sqrt{n}}A(z)\times\cos\left[\ln(n)\theta(z)+\delta(z)\right] \quad (76)$$

where the three unknown functions $A(z)$, $\theta(z)$ and $\delta(z)$ also depend on the polynomial parameters $\{\sigma, \mu, \nu\}$. As before, the asymptotics of this polynomial is not known like (54) hence the physical features (phase shift and energy spectrum) of the corresponding quantum mechanical system could be obtained analytically. However, when we use the Jacobi base with a specific space transformation we often get the recursion relation (74) with the corresponding specific potential function and numerically the eigenvalues hence the plot of the wavefunction using (16) and (12). However, the conventional energy spectrum of the potentials associated with this polynomial gives the energy spectrum in (76) as

$$z_n^2 = -2\left(n + \frac{\nu+1}{2} - \sqrt{-\sigma}\right)^2 \tag{77}$$

where $n = 0, 1, 2, ..., N$ and $N$ is the largest integer less than or equal to $\frac{\nu+1}{2} - \sqrt{-\sigma}$. A further insights, therefore shows that the scattering amplitude $A(z)$ in the generalized asymptotic (76) occurs at $i\sqrt{-z_n^2}$. The known scattering states of quantum systems associated with this polynomial give phase shift

$$\delta(z) = \arg\Gamma\left(i\sqrt{2}z\right) - \arg\Gamma\left(\frac{\nu+1}{2} - \sqrt{-\sigma} + \frac{i}{\sqrt{2}}z\right) - \arg\Gamma\left(\frac{\nu+1}{2} + \sqrt{-\sigma} + \frac{i}{\sqrt{2}}z\right) \tag{78}$$

and suggesting that scattering amplitude is proportional to

$$A(z) \propto \left|\Gamma\left(i\sqrt{2}z\right)\right| \bigg/ \left|\Gamma\left(\frac{\nu+1}{2} - \sqrt{-\sigma} + \frac{i}{\sqrt{2}}z\right)\Gamma\left(\frac{\nu+1}{2} + \sqrt{-\sigma} + \frac{i}{\sqrt{2}}z\right)\right| \tag{79}$$

With these, it implies that the continuous weight function for this polynomial takes the following form

$$\rho(z) \propto \frac{1}{A^2(z)} \Box \left|\Gamma\left(\frac{\nu+1}{2} - \sqrt{-\sigma} + \frac{i}{\sqrt{2}}z\right)\Gamma\left(\frac{\nu+1}{2} + \sqrt{-\sigma} + \frac{i}{\sqrt{2}}z\right) \bigg/ \Gamma\left(i\sqrt{2}z\right)\right|^2 \tag{80}$$

Finally, for negative $\sigma$ the orthogonality relation takes the form

$$\int_0^\infty \rho(z) G_n^{(\mu,\nu)}(z^2;\sigma) G_m^{(\mu,\nu)}(z^2;\sigma) dz + \sum_{k=0}^N \omega_k G_n^{(\mu,\nu)}(z_k^2;\sigma) G_m^{(\mu,\nu)}(z_k^2;\sigma) = \delta_{n,m} \tag{81}$$

where $G_n^{(\mu,\nu)}(z^2;\sigma) = A_n \bar{G}_n^{(\mu,\nu)}(z^2;\sigma)$ is the orthonormal version of $\bar{G}_n^{(\mu,\nu)}(z^2;\sigma)$ and $\omega_k$ is the normalized discrete weight function. And for positive $\sigma$, the summation sign will fade out. We observe that this polynomial has a discrete form with finite spectrum when $z \to iz_k$ in (74) where $k, n = 0, 1, ..., N$ and $\sigma$ is some proper function of $N$.

Concluding this section, we have given the orthogonal polynomials we usually encountered when using the Laguerre and Jacobi bases. For the Laguerre bases, the orthogonal polynomials are conventional ones. While using the Jacobi bases, we get new orthogonal polynomials not known in mathematics literature to the best of our knowledge and we encourage researchers in this field to look into getting the properties of these polynomials. However, using TRA, we show how (71) or (74) are gotten (with their discrete forms as the case maybe) when using the Jacobi bases.

**5. Application of TRA to conventional and new quantum system**

Now Table 2 contains some of the conventional potential functions solvable in TRA. Also we give the coordinate space transformation that can be used in TRA to get each potential function. We advise readers to follow each of the examples below to get similar results and also provide the coordinate space transformation for Rosen- Morse I and Rosen- Morse II.

The first illustration is the Spherical oscillator where we use the Laguerre basis element. Conventionally, this is three dimensional quantum systems. First, we use the radial time independent Schrodinger wave equation for structureless particle in a spherically symmetric potential $V(r) = \frac{1}{2}\omega^4 r^2$ which reads as

$$(H - E)|\psi\rangle = \left[-\frac{1}{2}\frac{d^2}{dr^2} + \frac{\ell(\ell+1)}{2r^2} + V(r) - E\right]|\psi\rangle = 0 \tag{82}$$

where $\ell$ is the angular momentum quantum number and $r$ is the radial coordinate in three dimensions. In TRA, there is no need to substitute for $V(r)$ in (82). We therefore use the space coordinate transformation $y = (\lambda r)^2$. Now using (27), then (82) becomes

$$(H-E)|\psi\rangle = -\frac{1}{2}\left[(y')^2\frac{d^2}{dy^2} + y''\frac{d}{dy} - \frac{\ell(\ell+1)}{r^2} - 2V(y) + 2E\right]|\psi\rangle \qquad (83)$$

since $y = (\lambda r)^2$ implies $y' = 2\lambda^2 r = 2\lambda\sqrt{y}$ and $y'' = 2\lambda^2$. Then (83) becomes

$$(H-E)|\psi\rangle = -\frac{1}{2}\left[(y')^2\frac{d^2}{dy^2} + y''\frac{d}{dy} - \frac{\ell(\ell+1)}{r^2} - 2V(y) + 2E\right]|\psi\rangle$$

$$= -\frac{1}{2}\left[4\lambda^2 y\frac{d^2}{dy^2} + 2\lambda^2\frac{d}{dy} - \frac{\ell(\ell+1)\lambda^2}{y} - 2V(y) + 2E\right]|\psi\rangle$$

$$= -\frac{4\lambda^2}{2}\left[y\frac{d^2}{dy^2} + \frac{1}{2}\frac{d}{dy} - \frac{\ell(\ell+1)}{4y} - \frac{1}{2\lambda^2}(V(y) - E)\right]|\psi\rangle$$

$$= -2\lambda^2\left[y\frac{d^2}{dy^2} + \frac{1}{2}\frac{d}{dy} - \frac{\ell(\ell+1)}{4y} - \frac{1}{2\lambda^2}(V(y) - E)\right]|\psi\rangle$$

$$\frac{1}{2\lambda^2}(H-E)|\psi\rangle = \left[-y\frac{d^2}{dy^2} - \frac{1}{2}\frac{d}{dy} + \frac{\ell(\ell+1)/4}{y} + \frac{1}{2\lambda^2}(V(y) - E)\right]|\psi\rangle \qquad (84)$$

Table 2: Some exactly Solvable Potential functions in Quantum mechanics with their corresponding coordinate space transformation in TRA. Here we have redefined the physical parameter and used conventional units $\hbar = m = 1$ or $\hbar = 2m = 1$.

| S/N | Name of Potential Function | Potential function $V(x)$ | Space Transformation $y(x)$ | Basis element $\phi(x)$ |
|---|---|---|---|---|
| 1 | Spherical oscillator | $\frac{1}{2}\bar{\omega}^4 r^2$ <br> $0 \leq r < \infty$ | $x = (\lambda r)^2$ | Laguerre Basis |
| 2 | Coulomb | $-\frac{Z^2}{x} + \frac{Z^4}{2}$ <br> $0 \leq x < \infty$ | $y(x) = \lambda x$ <br> $0 \leq x < \infty$ | Laguerre Basis |
| 3 | Morse | $V_0 + V_1 e^{-2\alpha x} - V_2 e^{-\alpha x}$ <br> $-\infty < x < \infty$ | $y = e^{\lambda x}$ <br> $-\infty < x < \infty$ | Laguerre Basis |
| 4 | Trigonometric Scarf potential | $V_0 + V_1 \cosec^2 \alpha x$ <br> $-V_2 \cot \alpha x \cos ec\alpha x$ <br> $0 \leq \alpha x \leq \pi$ | $y = \sin(\pi x/L)$ <br> $-L/2 \leq x \leq L/2$, <br> $\lambda = \pi/L$ | Jacobi Basis |
| 5 | Hyperbolic Scarf potential | $V_0 + V_1 \sech^2 \alpha x$ <br> $+V_2 \sech \alpha x \tanh \alpha x$ <br> $-\infty < x < \infty$ | $y(x) = 2\tanh^2(\lambda x) - 1$ <br> $0 \leq x < \infty$ | Jacobi Basis |
| 6 | Rosen- Morse I | $V_0 + V_1 \tan \alpha x + V_2 \sec^2 \alpha x$ <br> $-\infty < x < \infty$ | | Jacobi Basis |
| 7 | Rosen - Morse II | $V_0 + V_1 \tanh \alpha x - V_2 \sech^2 \alpha x$ <br> $-\infty < x < \infty$ | $y(x) = \tanh \lambda x$ | Jacobi Basis |
| 8 | Rosen – Morse II | $V_0 + V_1 \cosech^2 \alpha x$ <br> $-V_2 \coth \alpha x \times \cosech \alpha x$ <br> $0 \leq x < \infty$ | | Jacobi Basis |
| 9 | Hyperbolic Eckart | $V_0 - V_1 \coth \alpha x + V_2 \cosech^2 \alpha x$ <br> $0 \leq x < \infty$ | $y(x) = 1 - 2e^{-\lambda x}$ <br> $0 \leq x < \infty$ | Jacobi Basis |
| | | | | |

| 10 | Poschl-Teller I | $-V_0 + V_1 \sec^2 \alpha x + V_2 \csc^2 \alpha x$ $0 \le \alpha x \le \dfrac{\pi}{2}$ | $y(x) = 2\sin^2(\pi x/2L) - 1$ $0 \le x \le L$ | Jacobi Basis |
|---|---|---|---|---|
| 11 | Poschl-Teller II | $V_0 - V_1 \operatorname{sech}^2 \alpha x + V_2 \operatorname{csch}^2 \alpha x$ $0 \le x < \infty$ | $y(x) = 2\tanh^2(\lambda x) - 1$ $0 \le x < \infty$ | Jacobi Basis |

Writing in similar form to (45), we have

$$\frac{1}{2\lambda^2}\sum_{n=0}^{\infty} f_n(\varepsilon)\langle\phi_m(y)|(H-E)|\phi_n(y)\rangle = \sum_{n=0}^{\infty} f_n(\varepsilon)\langle\phi_m(y)|\left[-y\frac{d^2}{dy^2} - \frac{1}{2}\frac{d}{dy} + \frac{\ell(\ell+1)/4}{y} + \frac{1}{2\lambda^2}(V(y)-E)\right]|\phi_n(y)\rangle \quad (85)$$

and the wave operator matrix element is

$$J_{mn} = \frac{1}{2\lambda^2}\langle\phi_m(y)|(H-E)|\phi_n(y)\rangle = \langle\phi_m(y)|\left[-y\frac{d^2}{dy^2} - \frac{1}{2}\frac{d}{dy} + \frac{\ell(\ell+1)/4}{y} + \frac{1}{2\lambda^2}(V(y)-E)\right]|\phi_n(y)\rangle \quad (86)$$

with integral measure of $\int_0^\infty dr.... = \int_0^\infty \dfrac{dy}{2\lambda\sqrt{y}}....$ Hence the basis element is $|\phi_n(x)\rangle = A_n y^\alpha e^{-\beta y} L_n^v(y)$ and $A_n = \sqrt{2\lambda\Gamma(n+1)/\Gamma(n+v+1)}$. Since there is a space configuration transformation the basis element would be written as $|\phi_n(y)\rangle = A_n y^\alpha e^{-\beta y} L_n^v(y)$. Now, we follow a step by step procedure of evaluating (86) by starting from the R.H.S. So we have

$$\frac{1}{2\lambda^2}(H-E)|\phi_n(y)\rangle = \left[-y\frac{d^2}{dy^2} - \frac{1}{2}\frac{d}{dy} + \frac{\ell(\ell+1)/4}{y} + \frac{1}{2\lambda^2}(V(y)-E)\right]|\phi_n(y)\rangle \quad (87)$$

Since $\dfrac{d\phi_n}{dr} = \dfrac{d\phi_n}{dy}\times\dfrac{dy}{dr} = 2\lambda^2 r \times \dfrac{d\phi_n}{dy} = 2\lambda\sqrt{y}\dfrac{d\phi_n}{dy}$, implies that

$$\frac{d^2\phi_n}{dr^2} = 4\lambda^2 y\frac{d^2\phi_n}{dy^2} = 4\lambda^2 y\frac{d^2\phi_n}{dy^2} = 4\lambda^2 A_n y^{\alpha+1} e^{-\beta y}\left[\frac{d^2}{dy^2} + \left(\frac{2\alpha}{y} - 2\beta\right)\frac{d}{dy} - \frac{\alpha}{y^2} + \left(\frac{\alpha}{y} - \beta\right)^2\right] L_n^v(y) \quad (88)$$

where we have used (50). However, there will be no need for (88) in (87) since we had already transformed (82). So using (50) in (87), we have

$$\frac{1}{2\lambda^2}(H-E)|\phi_n(y)\rangle = \left[-y\left[\frac{d^2}{dy^2} + \left(\frac{2\alpha}{y} - 2\beta\right)\frac{d}{dy} - \frac{\alpha}{y^2} + \left(\frac{\alpha}{y} - \beta\right)^2\right] - \frac{1}{2}\frac{d}{dy} + \frac{\ell(\ell+1)/4}{y} + \frac{1}{2\lambda^2}(V(y)-E)\right] L_n^v(y) \quad (89)$$

$$= \left[-y\frac{d^2}{dy^2} - \left(2\alpha + 2y\beta + \frac{1}{2}\right)\frac{d}{dy} + \frac{\ell(\ell+1)/4 - \alpha(\alpha-1)}{y} - 2\alpha\beta + y\beta^2 + \frac{1}{2\lambda^2}(V(y)-E)\right] L_n^v(y)$$

Using (41) in (89), we have

$$\frac{1}{2\lambda^2}(H-E)|\phi_n(y)\rangle = \left[\left(v + \frac{1}{2} - 2\alpha - 2y\beta\right)\frac{d}{dy} + \frac{\ell(\ell+1)/4 - \alpha(\alpha-1)}{y} - 2\alpha\beta + y\beta^2 + n + \frac{1}{2\lambda^2}(V(y)-E)\right] L_n^v(y) \quad (90)$$

Furthermore, we use (42) in (90), we have

$$\frac{1}{2\lambda^2}(H-E)|\phi_n(y)\rangle = A_n y^\alpha e^{-\beta y}\left[\left(v + \frac{1}{2} - 2\alpha\right)\frac{n}{y} + \frac{\ell(\ell+1)/4 - \alpha(\alpha-1)}{y} - 2\alpha\beta + y\beta^2 + n + \frac{1}{2\lambda^2}(V(y)-E)\right] L_n^v(y) \quad (91)$$

$$- \left(v + \frac{1}{2} - 2\alpha\right)\frac{(n+v)}{y} L_{n-1}^v(y)$$

By using $A_n y^\alpha e^{-\beta y} L_{n-1}^v = \dfrac{A_n}{A_{n-1}}\phi_{n-1}$ in (91), we have

$$\frac{1}{2\lambda^2}(H-E)|\phi_n(y)\rangle = \left[-\left(v+\frac{1}{2}-2\alpha\right)\frac{n}{y}+\frac{\ell(\ell+1)/4-\alpha(\alpha-1)}{y}+\frac{1}{4}+2\alpha\beta-y\beta^2+n+\frac{1}{2\lambda^2}(V(y)-E)\right]\phi_n(y) \quad (92)$$

$$-\left(v+\frac{1}{2}-2\alpha\right)\frac{(n+v)}{y}\frac{A_n}{A_{n-1}}\phi_{n-1}$$

Now we take a critical look at the orthogonality relation (43) and the tridiagonal requirement, this limits the possibilities to either one of the following two:

$$(a) \quad \beta=\frac{1}{2}, v=2\alpha-\frac{1}{2}, \quad 2\alpha=\ell+1, \quad \text{and} \quad V(y)=A+By=\frac{1}{2}\omega^4 r^2 \quad (93a)$$

The potential function here easily gives $V(r)=\frac{1}{2}\omega^4 r^2$, where $\omega$ is the oscillator frequency.

$$(b) \beta=\frac{1}{2}, \quad v=2\alpha-\frac{3}{2}, \quad \text{and} \quad V(r)=\frac{1}{2}\lambda^4 r^2 + B/2r^2 \quad (93b)$$

The potential function here easily gives $V(r)=\frac{1}{2}\lambda^4 r^2 + B/2r^2$, where $B$ is a centripetal potential barrier parameter. The potential in the first condition was gotten by taking $A=0$, which gives $V(y)=By$, implies that $V(r)=B\lambda^2 r^2$ by taking $y=(\lambda r)^2$ and finally $V(r)=\frac{1}{2}\omega^4 r^2$ where $B=\frac{\omega^4}{2\lambda^2}$. Knowing these conditions is based on detailed study of equation (92) and the orthogonality property of the Laguerre polynomial. So, from equation (93) we can see the advantage of TRA in providing two possible forms of the spherical potential function. Now, we work with (93a), with necessary calculations, and leave (93b) as an exercise. Applying (93a) in (92), while reserving the expression for $V(y)$. Using (93a) in (92), we have

$$\frac{1}{2\lambda^2}(H-E)|\phi_n\rangle = \left[\frac{4(\alpha-1/4)^2-(\ell+1/2)^2}{4y}+n+\alpha+\frac{1}{4}-\frac{y}{4}+\frac{1}{2\lambda^2}(V-E)\right]|\phi_n\rangle \quad (94)$$

$$= \left[\frac{2n+v+1}{2}-\frac{y}{4}+\frac{1}{2\lambda^2}(V-E)\right]|\phi_n\rangle$$

Now writing in form of (46)

$$J_{mn}=\frac{1}{2\lambda^2}\langle\phi_m|(H-E)|\phi_n\rangle = \frac{1}{2\lambda^2}A_m A_n \int_0^\infty y^{2\alpha}e^{-2\beta y}L_n^v(y)\left[\frac{2n+v+1}{2}-\frac{y}{4}+\frac{1}{2\lambda^2}(V-E)\right]L_m^v(y)\cdot\frac{dy}{2\lambda\sqrt{y}} \quad (95)$$

$$= \frac{1}{2\lambda^2}\cdot\sqrt{\frac{2\lambda\Gamma(n+1)}{\Gamma(n+v+1)}}\cdot\sqrt{\frac{2\lambda\Gamma(m+1)}{\Gamma(m+v+1)}}\int_0^\infty y^{2\alpha-1/2}e^{-2\beta y}L_n^v(y)\left[\frac{2n+v+1}{2}-\frac{y}{4}+\frac{1}{2\lambda^2}(V-E)\right]L_m^v(y)\cdot\frac{dy}{2\lambda}$$

$$= \frac{1}{2\lambda^2}\sqrt{\frac{\Gamma(n+1)}{\Gamma(n+v+1)}}\cdot\sqrt{\frac{\Gamma(m+1)}{\Gamma(m+v+1)}}\int_0^\infty y^v e^{-y}L_n^v(y)\left[\frac{2n+v+1}{2}-\frac{y}{4}+\frac{1}{2\lambda^2}(V-E)\right]L_m^v(y)dy$$

Using the orthogonality property of Laguerre polynomial, the first expression in bracket under the integral gives

$$\frac{1}{2\lambda^2}\sqrt{\frac{\Gamma(n+1)}{\Gamma(n+v+1)}}\cdot\sqrt{\frac{\Gamma(m+1)}{\Gamma(m+v+1)}}\cdot\frac{2n+v+1}{2}\int_0^\infty y^v e^{-y}L_n^v(y)L_m^v(y)dy \quad (95a)$$

$$= \frac{1}{4\lambda^2}(2n+v+1)\cdot\frac{\Gamma(n+1)}{\Gamma(n+v+1)}\cdot\frac{\Gamma(n+v+1)}{\Gamma(n+1)}\delta_{mn}$$

$$= \frac{1}{4\lambda^2}(2n+v+1)\delta_{mn}$$

Similarly, the second expression will give

$$\frac{1}{2\lambda^2}\sqrt{\frac{\Gamma(n+1)}{\Gamma(n+v+1)}}\cdot\sqrt{\frac{\Gamma(m+1)}{\Gamma(m+v+1)}}\int_0^\infty y^{2\alpha-1/2}e^{-2\beta y}L_n^v(y)\left[-\frac{y}{4}\right]L_m^v(y)dy \quad (95b)$$

$$= \left\{\left\{-\frac{1}{8\lambda^2}(2n+v+1)\delta_{mn}+\frac{1}{8\lambda^2}\sqrt{n(n+v)}\delta_{n,m+1}+\frac{1}{8\lambda^2}\sqrt{(n+1)(n+v+1)}\delta_{n,m-1}\right\}\right.$$
~ 18 ~

using the recurrence relation of the Laguerre polynomial $yL_n^v(y) = (2n+v+1)L_n^v(y) - (n+v)L_{n-1}^v(y) - (n+1)L_{n+1}^v(y)$.

Now, for the potential function term since $V(r) = \frac{1}{2}\omega^4 r^2$ which can also be written as $V(y) = \left(-\frac{2\omega^4}{\lambda^2}\right)\left(-\frac{y}{4}\right)$, therefore we have

$$\frac{1}{2\lambda^2} \sqrt{\frac{\Gamma(n+1)}{\Gamma(n+v+1)}} \cdot \sqrt{\frac{\Gamma(m+1)}{\Gamma(m+v+1)}} \cdot \int_0^\infty y^v e^{-y} L_n^v(y) \left[\frac{V}{2\lambda^2}\right] L_m^v(y) dy \qquad (95c)$$

$$= \frac{-2\omega^4}{\lambda^2}\left\{-\frac{1}{8\lambda^2}\delta_{mn} + \frac{1}{8\lambda^2}\sqrt{n(n+v)}\delta_{n,m+1} + \frac{1}{8\lambda^2}\sqrt{(n+1)(n+v+1)}\delta_{n,m-1}\right\}$$

and finally, we have the last term has $-\frac{E}{4\lambda^4}\delta_{mn}$. Using (95a-c) in (95) with necessary simplifications, we have

$$\frac{2}{\lambda^2}\langle\phi_m|(H-E)|\phi_n\rangle = \frac{1}{\lambda^2}\left\{\begin{array}{l}\left[(2n+v+1)\left(\frac{1}{2}+\frac{\omega^4}{\lambda^2}\right) - \frac{E}{\lambda^2}\right]\delta_{nm} + \left(\frac{1}{2} - \frac{\omega^4}{\lambda^2}\right) \\ \left[\sqrt{n(n+v)}\delta_{n,m+1} + \sqrt{(n+1)(n+v+1)}\delta_{n,m-1}\right]\end{array}\right\} \qquad (96)$$

and by using (45), we have the recurrence relation

$$(2n+v+1)\left(\frac{\omega^4}{\lambda^2} + \frac{1}{2}\right)f_n = \frac{E}{\lambda^2}f_n + \left(\frac{\omega^4}{\lambda^2} - \frac{1}{2}\right)\left[\sqrt{n(n+v)}f_{n-1} + \sqrt{(n+1)(n+v+1)}f_{n+1}\right] \qquad (97)$$

$$(2n+v+1)\left(\frac{z+1}{z-1}\right)f_n = \frac{E/\lambda^2}{z-1}f_n + \sqrt{n(n+v)}f_{n-1} + \sqrt{(n+1)(n+v+1)}f_{n+1}$$

where $z = 2\omega^4/\lambda^2$. In order to find the expansion coefficient $f_n(\varepsilon)$ of wavefunction from (97), we compare it to (53), the recurrence relation for Meixner Pollaczek polynomial

$$2y\sin\theta P_n^\mu(y,\theta) = -\left[(2n+2\mu)\cos\theta\right]P_n^\mu(y,\theta) + \sqrt{n(n+2\mu-1)}P_{n-1}^\mu(y,\theta) + \sqrt{(n+1)(n+2\mu)}P_{n+1}^\mu(y,\theta) \qquad (98)$$

Now if we take $2\mu = v+1$, $y\sin\theta = -\frac{E/\lambda^2}{z-1}$, and $\cos\theta = \frac{z+1}{z-1}$. As $z \to \infty$, then $\cos\theta$ tends towards -1 and 1 which is hyperbolic function $\cosh\theta$. So we rewrite (98) in hyperbolic form (Hyperbolic Meixner Pollaczek) in order to make it same as (97). We take $\theta \to i\theta$, then $\sin\theta = i\sinh\theta$, $\cos\theta = \cosh\theta$, and (98), becomes

$$i(2y\sinh\theta)f_n^\mu(y,\theta) = -\left([2n+2\mu]\right)\cosh\theta f_n^\mu(y,\theta) + \sqrt{n(n+v)}f_{n-1}^\mu(y,\theta) + \sqrt{(n+1)(n+v+1)}f_{n+1}^\mu(y,\theta) \qquad (99)$$

which is same as (97). So the expansion coefficient $f_n^\mu(y,\theta)$ of the wavefunction is gotten from (51) by taking $\theta \to i\theta$ to give $f_n^\mu(y,\theta) = \sqrt{\frac{(2\mu)_n}{n!}}e^{-n\theta}{}_2F_1\left(\begin{array}{c}-n,\mu+iy\\2\mu\end{array}\bigg|1-e^{-2\theta}\right)$. And since $\cos\theta = \cosh\theta = \frac{z+1}{z-1}$, then $\sinh\theta = \frac{z\sqrt{2}}{z-1}$, therefore

$y\sinh\theta = -\frac{E/\lambda^2}{z-1}$ will give $y = -E/2\sqrt{2}\lambda\omega^2$. The discrete energy spectrum of the quantum system is gotten from the zero of scattering amplitude (55) of the Meixner Pollaczek polynomial. For the hyperbolic of this polynomial, we have $y^2 = (n+\mu)^2$ which gives $\frac{E^2}{8\lambda^2\omega^4} = \left(n + \frac{v+1}{2}\right)^2$ and finally we have the energy spectrum as $E = \omega^2\left(2n+\ell+\frac{3}{2}\right)$. It can be seen that spherically symmetric potential $V(r) = \frac{1}{2}\omega^4 r^2$ was not used in (82) yet was recovered in (93) because we used the right coordinate space transformation. This is the beauty of TRA.

Now, we present another detailed example for Jacobi basis, where we solved a new three parameters potential function using TRA. This potential function is given as:

$$V(x) = \frac{1}{e^{\lambda x} - 1}\left[V_0 + V_1\left(1 - 2e^{-\lambda x}\right) + \frac{V_R}{1 - e^{-\lambda x}}\right] \qquad (100)$$

for $0 \leq x \leq \infty$, $V_i$ are real parameters such that $V_R$ is greater than or equal to zero. Fig 1 is the plot of the potential function where $V_R$ is vary (keeping $V_0$ and $V_1$ constant) and varying $V_0$ and $V_1$ (keeping $V_R$ constant). However, when deformed, this potential can be compared to a three – parameter potential in

$$V(x) = V_1 \frac{e^{-\lambda x} - \gamma}{e^{\lambda x} - 1} \tag{101}$$

where $V_1$ is the potential strength and the range parameter $\lambda$ is positive with inverse length unit. The dimensionless parameter $\gamma$ is in the open range $0 < \gamma < 1$. Graphically, this potential function cross the $x$ –axis at $x_0 = -\ln \gamma / \lambda$ and then reaches a local extremum value of $V(x_1) = -V_0 \left(1 - \sqrt{1-\gamma}\right)^2$ at $x_1 = -\frac{1}{\lambda} \ln\left(1 - \sqrt{1-\gamma}\right)$. We observed that (100) and (101) are short range potentials with $1/x$ singularity at the origin. Simple algebraic manipulation in (100) will easily produce (101). Note (100) is same as

$$V(x) = \frac{V_0}{e^{\lambda x} - 1} + \frac{V_1 \left(1 - 2e^{-\lambda x}\right)}{e^{\lambda x} - 1} + \frac{V_R e^{\lambda x}}{\left(e^{\lambda x} - 1\right)^2} \tag{102}$$

$$= V_1 \frac{\gamma - e^{-\lambda x}}{e^{\lambda x} - 1} + \frac{V_R e^{\lambda x}}{\left(e^{\lambda x} - 1\right)^2}$$

when $V_R = 0$, such that we defined a dimensionless ratio $\gamma = \frac{V_0 + V_1}{2V_1}$ for $V_1 \neq 0$; then we easily get (101). Detailed study of these potentials; show that they can be used practically as an appropriate model for the interaction of an electron with extended molecules whose electron cloud is congregated near the centre of the molecules. Also, there is a resemblance between these potentials and attractive Coulomb potential for nonzero angular momentum at short distances hence we expect them to finite bound states. However, we are more interested in the generalized potential given in (100) where none of the potential parameters is zero in order to obtain the energy spectrum and wavefunction.

Now we make the coordinate transformation $y(x) = 1 - 2e^{-\lambda x}$, then the Schrödinger wave equation in the new configuration space becomes

$$(H - E)|\psi\rangle = -\frac{1}{2}\left[(y')^2 \frac{d^2}{dy^2} + y'' \frac{d}{dy} - 2V(y) + 2E\right]|\psi\rangle = 0, \tag{103}$$

where the prime stands for the derivative with respect to $x$ and we adopted the atomic units $\hbar = m = 1$. With $y \in [-1, +1]$, we can choose the following square integrable functions in the new configuration space with coordinate $y$ as basis elements for the expansion of the wavefunction

$$\phi_n(y) = A_n (1 - y)^\alpha (1 + y)^\beta P_n^{(\mu,\nu)}(y) \tag{104}$$

where $P_n^{(\mu,\nu)}(x)$ is the Jacobi polynomial of degree $n = 0, 1, 2,..$ in $y$, the parameters $\mu$ and $\nu$ are larger than $-1$ and $A_n = \sqrt{\frac{2n+\mu+\nu+1}{2^{\mu+\nu+1}} \frac{\Gamma(n+1)\Gamma(n+\mu+\nu+1)}{\Gamma(n+\nu+1)\Gamma(n+\mu+1)}}$. For simplicity, we rewrite (103) as

$$(H - E)|\psi\rangle = -\frac{1}{2} \frac{(y')^2}{(1-y^2)}\left[(1-y^2)\frac{d^2}{dy^2} + \frac{y''}{(y')^2}(1-y^2)\frac{d}{dy} + 2\frac{(1-y^2)}{(y')^2}[E - V]\right]|\psi\rangle = 0 \tag{105}$$

We applied the wave operator on the basis element to get

$$(H - E)|\phi_n\rangle = J|\phi_n\rangle = -\frac{\lambda^2}{2} \frac{(1-y)}{(1+y)}\left[(1-y^2)\frac{d^2}{dy^2} - (1+y)\frac{d}{dy} + \frac{2(1+y)E}{\lambda^2(1-y)} - \frac{2V_0}{\lambda^2} - \frac{2V_1 y}{\lambda^2} - \frac{4V_R}{\lambda^2(1+y)}\right]|\phi_n\rangle = 0 \tag{106}$$

where we had used $y' = \lambda(1-y)$, $\frac{y''}{(y')^2} = -1/(1-y)$, and the potential function in new configuration space as

$$V(y) = \frac{(1-y)}{(1+y)}\left[V_0 + V_1 y + \frac{2V_R}{(1+y)}\right]. \text{ Using } u_i = \left(\frac{2}{\lambda^2}\right)V_i \text{ and } \varepsilon = \left(\frac{2}{\lambda^2}\right)E \text{ in (106) we have}$$

$$\frac{2}{\lambda^2} J |\phi_n\rangle = -\frac{(1-y)}{(1+y)}\left((1-y^2)\frac{d^2}{dy^2}-(1+y)\frac{d}{dy}-(\varepsilon+u_0)-u_1 y+\frac{2\varepsilon}{(1-y)}-\frac{2u_R}{(1+y)}\right)|\phi_n\rangle = 0 \qquad (107)$$

Now, the boundary conditions and square integrability (with respect to the integral measure $dx$) dictate that the matrix wave operator becomes

$$\langle \phi_n | F[y] | \phi_m \rangle = -A_n A_m \int_{-1}^{+1} (1-y)^{2\alpha}(1+y)^{2\beta}\frac{(1-y)}{(1+y)}[F(y)] P_n^{(\mu,v)}(y) P_m^{(\mu,v)}(y) \frac{dy}{\lambda(1-y)}$$

$$= \frac{-A_n A_m}{\lambda}\int_{-1}^{+1}(1-y)^{2\alpha}(1+y)^{2\beta-1}[F(y)] P_n^{(\mu,v)}(y) P_m^{(\mu,v)}(y) dy \qquad (108)$$

where $F(y) = (1-y^2)\frac{d^2}{dy^2}-(1+y)\frac{d}{dy}-(\varepsilon+u_0)-u_1 y+\frac{2\varepsilon}{(1-y)}-\frac{2u_R}{(1+y)}$. Hence we have $2\alpha = \mu$ and $2\beta = v+1$. As a result of these conditions with the fact that $y' = \lambda(1-y)$, the normalization constant will be $A_n = \sqrt{\frac{2n+\mu+v+1}{2^{\mu+v+1}}\frac{\Gamma(n+1)\Gamma(n+\mu+v+1)}{\Gamma(n+v+1)\Gamma(n+\mu+1)}}$.

Using the first (33) and second derivatives (34) of the basis element with respect to $y$ in (107) gives

$$\frac{2}{\lambda^2}J|\phi_n\rangle = -A_n(1-y)^\alpha(1+y)^\beta \begin{pmatrix} (1-y^2)\dfrac{d^2}{dy^2}+[2\beta(1-y)-(2\alpha+1)(1+y)]\dfrac{d}{dy}+\dfrac{1}{(1+y)}(2\beta(\beta-1)-2u_R) \\ -\dfrac{1}{(1-y)}(2\alpha^2+2\varepsilon)-(\alpha^2+\beta^2)-(\varepsilon+u_0)-u_1 y \end{pmatrix} P_n^{(\mu,v)}(y) = 0 \qquad (109)$$

The second order differential equation of the Jacobi Polynomial $P_n^{(\mu,v)}(y)$ is

$$(1-y^2)\frac{d^2}{dy^2}P_n^{(\mu,v)}(y) = \left\{[(\mu+v+2)y+\mu-v]\frac{d}{dy}-n(n+\mu+v+1)\right\}\frac{d}{dy} \qquad (110)$$

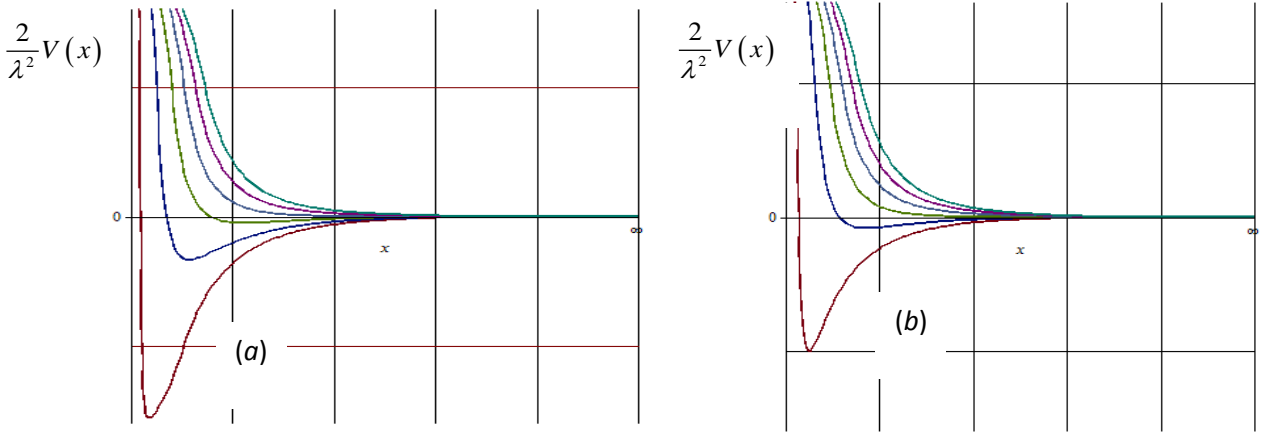

**Fig 1**. Plot of the potential function given by (1) with $\lambda = 1$. (a) is obtained by vary $u_r = 1,..,11$ (in step of 2) while keeping $u_1 = -5$ and $u_0 = -2$. (b) is obtained by $u_1 = -3$ and $u_0 = -2$ while $u_r = 1,..,11$ (in steps of 2). It is observed that further variations in $u_1$ and $u_0$ will result in an upper shift of the potential function in positive $xy$ plane.

Therefore equation (109) becomes

$$\frac{2}{\lambda^2}J|\phi_n\rangle = -A_n(1-y)^\alpha(1+y)^\beta \begin{pmatrix} [\mu-v+y(\mu+v+2)]\dfrac{d}{dy}+[2\beta(1-y)-(2\alpha+1)(1+y)]\dfrac{d}{dy}+\dfrac{1}{(1+y)}(2\beta(\beta-1)-2u_R) \\ +\dfrac{1}{(1-y)}(2\alpha^2+2\varepsilon)-(\alpha^2+\beta^2)-(\varepsilon+u_0)-u_1 y-n(n+\mu+v+1) \end{pmatrix} P_n^{(\mu,v)}(y)$$

$$= -A_n (1-y)^{\mu/2} (1+y)^{\nu+1/2} \left( \frac{1}{2(1+y)}(\nu^2 - 1 - 4u_R) + \frac{1}{2(1-y)}(\mu^2 + 4\varepsilon) - (\varepsilon + u_0) - u_1 y - \left(n + \frac{\mu+\nu+1}{2}\right)^2 \right) P_n^{(\mu,\nu)}(y) \tag{111}$$

Since the matrix representation of the wave operator is required to be tridiagonal and symmetric, in line with the recursion relation of the Jacobi polynomial and its orthogonality; we eliminate the two non-linear terms in (111). Hence the basis parameters must be chosen as follows:

$$\nu^2 = 1 + 4u_R \quad \text{and} \quad \mu^2 = -4\varepsilon \tag{112}$$

It is explicit here that the solution of this problem will give negative energy $\mu^2 = -8E/\lambda^2$ and the potential parameter $V_R$ ( $u_R \geq -1/4$ ) should be greater than or equal to zero. Now equation (111) becomes

$$\frac{2}{\lambda^2} J |\phi_n\rangle = -A_n (1-y)^{\mu/2} (1+y)^{\nu+1/2} \left( -(\varepsilon + u_0) - u_1 y - \left(n + \frac{\mu+\nu+1}{2}\right)^2 \right) P_n^{(\mu,\nu)}(y) \tag{113}$$

Using the three term recursion relation of the Jacobi polynomial and their orthogonality property,

$A_n A_m \int_{-1}^{+1} (1-y)^{\mu} (1+y)^{\nu} P_n^{(\mu,\nu)}(y) P_m^{(\mu,\nu)}(y) dy = \delta_{nm}$, in (113) after operating $\langle \phi_m |$ on the L.H.S of it; we have the tridiagonal and symmetric representation of the wave operator as

$$\frac{2}{\lambda^2} J_{nm} = \left[ \left(n + \frac{\mu+\nu+1}{2}\right)^2 + (\varepsilon + u_0) + u_1 C_n \right] \delta_{nm} + u_1 \left( D_{n-1} \delta_{n,m+1} + D_n \delta_{n,m-1} \right) \tag{114}$$

where $C_n = \frac{\nu^2 - \mu^2}{(2n+\mu+\nu)(2n+\mu+\nu+2)}$ and $D_n = \frac{2}{2n+\mu+\nu+2} \sqrt{\frac{(n+1)(n+\mu+1)(n+\nu+1)(n+\mu+\nu+1)}{(2n+\mu+\nu+1)(2n+\mu+\nu+3)}}$

Hence, we can now write the matrix wave equation as $\langle \phi_n | J | \psi \rangle = \sum_m \langle \phi_n | J | \phi_m \rangle f_m = \sum_m J_{nm} f_m = 0$ which give the three-term recursion relation for the expansion coefficient of the wave function as

$$-(\varepsilon + u_0) f_n = \left[ \left(n + \frac{\mu+\nu+1}{2}\right)^2 + u_1 C_n \right] f_n + u_1 \left( D_{n-1} f_{n-1} + D_n f_{n+1} \right) \tag{115}$$

writing $G_n(E) = G_0(E) f_0(\varepsilon)$, will make $f_0 = 1$ and $f_1 = \frac{(z-a_0) f_0}{b_0}$ where $z = -(\varepsilon + u_0)$, $a_0 = \left(\frac{\mu+\nu+1}{2}\right)^2 + u_1 C_0$, and $b_0 = u_1 D_0$. This relation is valid for $n = 1,2,3,\ldots$.

This is a new polynomial that we discovered recently and is not found in any mathematics literature. Hence, its analytic properties, that is, weight function, generating function, orthogonality, zero, etc. are yet to be known. Therefore we resolved to a numerical techniques to calculate the energy spectrum of the potential given in (100) for a given set of parameters. To calculate the energy spectrum, we obtain first the Hamiltonian matrix from the wave operator matrix (115) as $H = J|_{E=0}$. Then, the energy spectrum is calculated from the wave equation $H |\psi\rangle = E |\psi\rangle$ as the generalized eigenvalues $\{E\}$ of the matrix equation $\sum_m H_{n,m} f_m = E \sum_m \Omega_{n,m} f_m$; where $\Omega_{nm}$ is the overlap basis element given as

$$\Omega_{n,m} = \langle \phi_n | \phi_m \rangle = A_n A_m \int_{-1}^{1} (1-y)^{\mu} (1+y)^{\nu} P_n^{(\mu,\nu)}(y) P_m^{(\mu,\nu)}(y) \times \left[ (1-y)^{2\alpha-\mu-a} (1+y)^{2\beta-\nu-b} \right] dy \tag{116}$$

$$\equiv \langle n | (1-y)^{2\alpha-\mu-a} (1+y)^{2\beta-\nu-b} | m \rangle$$

Using the conditions for the parameters of the basis element $2\alpha = \mu$ and $2\beta = \nu + 1$ with $a = 1$ and $b = 0$; the overlap basis element becomes

$$\Omega_{nm} = \langle n | \frac{(1+y)}{(1-y)} | m \rangle = \frac{(1+C_n) \delta_{nm} + D_{n-1} \delta_{n,m+1} + D_n \delta_{n,m-1}}{(1-C_n) \delta_{nm} - D_{n-1} \delta_{n,m+1} - D_n \delta_{n,m-1}} \tag{117}$$

Table 3 is a list of the energy spectrum for a given set of values of the potential parameters and for basis size of $N = 20$. We show only significant decimal digits that do not change with any substantial increase in the basis size (e.g. from size 10 to 50). Also, variation (increase) in $N$; we observed a rapid convergence of these values with the size of the basis size

In Figure 2, we plot the bound state wavefunctions corresponding to the physical configuration and energy spectrum of Table 1. We calculate the $m^{th}$ bound state using the sum $\psi(E_m, x) \sim \sum_{n=0}^{N-1} P_n(\varepsilon_m)\phi_n(x)$, for some eigenvalues, where $N$ is some large enough integer. An energy resonance (scattering phase shift) phenomenon is observed with this potential function, which will observed when plotting the wavefunction using some of the eigenvalues in table 1.

**Table 3**: The finite energy spectrum for the potential function (1) with $u_0 = -6$, $u_1 = 10$, and $u_R = 2.5$. It was obtained by diagonalizing the Hamiltonian matrix for different basis size.

| $n$ | $\varepsilon_n$ | $n$ | $\varepsilon_n$ |
|---|---|---|---|
| 0 | 4126.9891447498 | 10 | 32.4394977694 |
| 1 | 1542.8903686294 | 11 | 22.4398560240 |
| 2 | 787.2186135745 | 12 | 15.2463797234 |
| 3 | 462.6214937613 | 13 | 10.1007245429 |
| 4 | 294.0660278716 | 14 | 6.4695140302 |
| 5 | 195.9554935304 | 15 | 3.9657920559 |
| 6 | 134.3972745542 | 16 | 2.2930653158 |
| 7 | 93.7323498172 | 17 | 0.0664830132 |
| 8 | 65.8910440926 | 18 | 0.4757637553 |
| 9 | 46.3583338365 | 19 | 1.1960489428 |

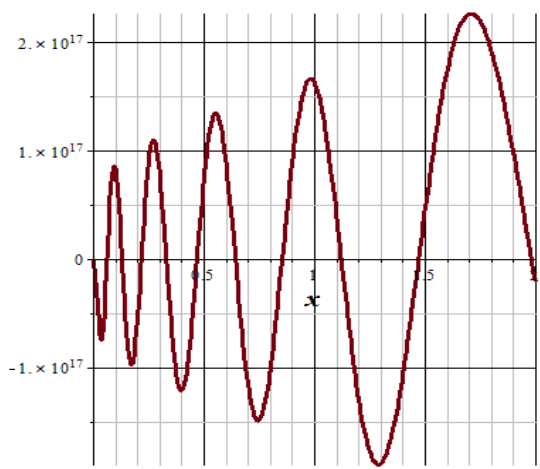
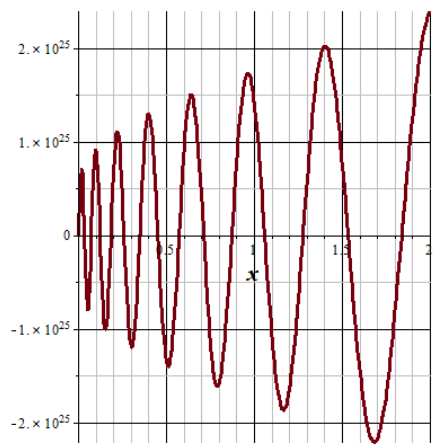

**FIG.2**. The graph of the wavefunction as we move down (however there coexist bound states and resonance) the table 1 with physical parameters: $\lambda = 1$, $u_0 = -6$, $u_1 = 10$, and $u_R = 2.5$.

## 6. Conclusion

In this paper, we introduced a pedagogical approach to TRA suitable for classroom teaching. The essence is to enable students of quantum mechanics class know how to use this algebraic method. This method (algebraic in nature) is encompassing then the conventional methods in solving the Schrödinger equation and allows for more realization of unknown potential functions which are yet to be associated with any physical systems. It has been shown that this approach has contributed greatly to the class of solvable quantum mechanical systems [8]

## Acknowledgement

The authors highly appreciate the support of the Saudi Centre for Theoretical Physics during the progress of this work. Specifically, this paper is dedicated to **Prof. A.D. Alhaidari** for his fatherly role and support in impacting knowledge. Allah in His infinite mercy will continue to bless him more. T. J. Taiwo specially appreciates his lovely wife – **Joan Ejiroghene Tunde** for her during the course of this work.

## Appendix A: Orthonormal Version of Orthogonal Polynomials

Orthonormal polynomials $\hat{P}_n(x)$ satisfy a three-term recursion relation

$$x\hat{P}_n(x) = a_n \hat{P}_n(x) + b_{n-1}\hat{P}_{n-1}(x) + b_n \hat{P}_{n+1}(x) \tag{A1}$$

and an orthogonal relation

$$\int_{x_-}^{x_+} \hat{\rho}(x)\hat{P}_n(x)\hat{P}_m(x)dx = \delta_{nm} \tag{A2}$$

where $\hat{\rho}(x)$ is the associated density function such that $\hat{P}_0(x)=1$ and $\int_{x_-}^{x_+}\hat{\rho}(x)=1$. On the other hand, standard orthogonal polynomials, $P_n(x)$, usually satisfy a three term recursion relation of the form

$$xP_n(x) = a_n P_n(x) + c_{n-1}P_{n-1}(x) + d_n P_{n+1}(x) \tag{A3}$$

an and orthogonality relation of the form

$$\int_{x_-}^{x_+} \rho(x)P_n(x)P_m(x)dx = \lambda_n \delta_{nm} \tag{A4}$$

where $\lambda_n > 0$ for all $n$. they also satisfy $P_0(x)=1$ giving $\int_{x_-}^{x_+}\rho(x) = \lambda_0$. Thus we conclude that $\hat{\rho}(x) = \rho(x)/\lambda_0$ and $\hat{P}_n(x) = \sqrt{\lambda_0/\lambda_n}P_n(x)$. Multiplying both sides of (A3) by $\sqrt{\lambda_0/\lambda_n}$, we obtain

$$x\hat{P}_n(x) = a_n \hat{P}_n(x) + c_{n-1}\sqrt{\frac{\lambda_{n-1}}{\lambda_n}}\hat{P}_{n-1}(x) + d_n\sqrt{\frac{\lambda_{n+1}}{\lambda_n}}\hat{P}_{n+1}(x) \tag{A5}$$

Now comparing this with (A1), we have $c_n = b_n\sqrt{\lambda_{n+1}/\lambda_n}$ and $d_n = b_n\sqrt{\lambda_n/\lambda_{n+1}}$. As an example, the standard Meixner-Pollaczek polynomial is defined as [Sect 1.7, 38]

$$P_n^\mu(x;\theta) = \frac{(2\mu)_n}{n!}e^{in\theta}\,{}_2F_1\left(\begin{matrix}-n,\mu+ix\\2\mu\end{matrix}\bigg|1-e^{-2i\theta}\right) \tag{A6}$$

with orthogonality

$$\frac{1}{2\pi}\int_{-\infty}^{\infty} e^{(2\theta-\pi)x}\left|\Gamma(\mu+ix)\right|^2 P_n^\mu(x;\theta)P_m^\mu(x;\theta)dx = \frac{\Gamma(n+2\mu)}{(2\sin\theta)^{2\mu}n!}\delta_{mn},\ \mu>0 \text{ and } 0<\theta<\pi \tag{A7}$$

and recurrence relation

$$x\sin\theta P_n^\mu(x;\theta) = -\left[(n+\mu)\cos\theta\right]P_n^\mu(x;\theta) + \frac{1}{2}(n+2\mu-1)P_{n-1}^\mu(x;\theta) + \frac{1}{2}(n+1)P_{n+1}^\mu(x;\theta) \tag{A8}$$

The normalized form this polynomial is gotten as follow: comparing (A4) to (A7), we have

$$\lambda_n = \frac{2\pi \Gamma(n+2\mu)}{(2\sin\theta)^{2\mu} n!} \tag{A9}$$

Therefore, normalized weight function is

$$\hat{\rho}(x) = \frac{\rho(x)}{\lambda_0} = \frac{e^{(2\theta-\pi)x}|\Gamma(\mu+ix)|^2}{\frac{2\pi\Gamma(2\mu)}{(2\sin\theta)^{2\mu}}} = \frac{1}{2\pi\Gamma(2\mu)}(2\sin\theta)^{2\mu} e^{(2\theta-\pi)x}|\Gamma(\mu+ix)|^2 \tag{A10}$$

Comparing (A5) to (A1), we have $b_n = c_n\sqrt{\lambda_n/\lambda_{n+1}}$, this easily gives

$$b_n = (n+2\mu)\times\sqrt{\frac{\Gamma(n+2\mu)}{(2\sin\theta)^{2\mu} n!}} \div \sqrt{\frac{\Gamma(n+2\mu+1)}{(2\sin\theta)^{2\mu}(n+1)!}} = \sqrt{(n+1)(n+2\mu)} \tag{A11}$$

So the normalized recurrence relation becomes

$$x\sin\theta P_n^\mu(x;\theta) = -\left[(n+\mu)\cos\theta\right]P_n^\mu(x;\theta) + \frac{1}{2}\sqrt{n(n+2\mu-1)}P_{n-1}^\mu(x;\theta) + \frac{1}{2}\sqrt{(n+1)(n+2\mu)}P_{n+1}^\mu(x;\theta) \tag{A12}$$

and (A6), becomes

$$P_n^\mu(x;\theta) = \sqrt{\frac{\Gamma(n+2\mu)}{\Gamma(2\mu)\Gamma(n+1)}} e^{in\theta} \,_2F_1\left(\begin{array}{c} -n, \mu+ix \\ 2\mu \end{array}\bigg| 1-e^{-2i\theta}\right) \tag{A13}$$

This is the normalized form of the Meixner Pollaczek polynomial. Also this procedure is universal for all orthogonal polynomials.

**Reference**

[1] A.D. Alhaidari, "*An extended class of $L^2$ - series solution of the wave equation,*" Ann. Phys **317,** 152 (2005)
[2] A.D. Alhaidari, " *Analytic solution of the wave equation for an eletron in the field of a molecule with an electric dipole moment*, " Ann. Phys, **323**, 1709 (2008)
[3] A.D. Alhaidari, "*Scattering and bound states for a class of non –central potentials*", J.Phys. A: Math. Theor.**38**. 3409 (2005)
[4] A.D. Alhaidari, "*A class of singular logarithmic potentials in a box with different skin thickness and wall interactions,*" Phys. Scr. **82**, 065008 (2010)
[5] A.D. Alhaidari, "*Exact solutions of Dirac and Schrodinger equations for a large class of power- law potentials at zero energy,*" Int. J. Mod. Phys. A **17,** 4551 (2002)
[6] P.C. Ojihe, "*SO(2,1) Lie algebra, the Jacobi matrix and scattering states of the Morse oscillator,*" J. Phys. A: Math. Gen **21**, 875 (1988)
[7] A.D. Alhaidari, " On the asymptotic solutions of the scattering problem", J. Phys. A: Math. Theor.**41** 175201 (2008)
[8] A.D. Alhaidari, *Solution of the nonrelativistic wave equation using the tridiagonal representation approach*, J. Math. Phys. **58,** 072104 (2017)
[9] R.W. Haymaker and L. Schlessinger, *The Pade Approximation in Theoretical Physics*, edited by G.A. Baker and J.L Gammel (Academic Press, New York, 1970)
[10] A.D. Alhaidari and H. Bahlouli, *Extending the class of solvable potentials I. The infinite potential well with a sinusoidal bottom,* J. Math. Phys. 49 (2008) 082102
[11] J.S. Geronimo and K. M. Case, *Scattering theory and polynomials orthogonal on the real line,* Tans. Am. Math. Soc**258**, 467 – 494 (1980)
[12] F.W.J. Oliver, *Asymptotic and Special Functions*, (Academic Press, New York, 1974)
[13] J.S. Geronimo, *A relation between the coefficient in the recurrence formula and the spectral Function for the orthogonal polynomials,* Tran. Am. Math. Soc. **260,** 65 -82 (1980)
[14] H .Kamada, Y. Koike, and W. Glockle, "*Complex energy method for scattering processes"*, Prog. Theor. Phys **109,** 869 (2003)
[15] N. Moiseyev, "*Quantum theory of resonances: calculating energies, widths and cross sections by complex scaling* ", Phys Rep. **302**, 212 (1998)
[16] R. Koekoek and R. Swarttouw, *The Askey –Scheme of Hypergeometric orthogonal polynomials and*

    *its q analogues,* Reports of the Faculty of Technical Mathematics and Informatics, Number 98 -17
    (Delft University of Technology, Delft, 1998)

[17] A. Bottino, A. Longoni, and A. Regge, "*Potential scattering for complex energy and angular momentum*
    II NuovoCimento (1995-1965) **23**, 954 (1962)